\documentclass[letterpaper, aps, prd, twocolumn, superscriptaddress, showpacs, nofootinbib]{revtex4}
\pdfoutput=1

\usepackage{graphicx}
\usepackage{bm}
\usepackage{dcolumn}
\usepackage{color}
\usepackage[sort&compress]{natbib}
\usepackage[latin9]{inputenc}
\usepackage{url}
\usepackage{subfigure}
\usepackage{float}
\usepackage{longtable}
\usepackage{amssymb}
\usepackage{amsmath}
\usepackage{amsfonts}
\begin{document}

\title{Inferring Astrophysical Parameters of Core-Collapse Supernovae from their Gravitational-Wave Emission}

\newcommand*{\ozgrav}{Centre for Astrophysics and Supercomputing, Swinburne University of Technology, Hawthorn, VIC 3122, Australia.}
\affiliation{\ozgrav}

\newcommand*{\monash}{School of Physics and Astronomy, Monash University, Clayton, VIC 3800 Australia.}
\affiliation{\monash}

\author{Jade Powell}  \affiliation{\ozgrav}
\author{Bernhard M\"uller} \affiliation{\monash}	

\date{\today}

\begin{abstract} 

Nearby core-collapse supernovae (CCSNe) are powerful multi-messenger sources for gravitational-wave, neutrino and electromagnetic telescopes as they emit gravitational waves in the ideal frequency band for ground based detectors. Once a CCSN gravitational-wave signal is detected, we will need to determine the parameters of the signal, and understand how those parameters relate to the source's explosion, progenitor and remnant properties. This is a challenge due to the stochastic nature of CCSN explosions, which is imprinted on their time series gravitational waveforms. In this paper, we perform Bayesian parameter estimation of CCSN signals using an asymmetric chirplet signal model to represent the dominant high-frequency mode observed in spectrograms of CCSN  gravitational-wave signals. We use design sensitivity Advanced LIGO noise and CCSN waveforms from four different hydrodynamical supernova simulations with a range of different progenitor stars. We determine how well our model can reconstruct time-frequency images of the emission modes, and show how well we can determine parameters of the signal such as the frequency, amplitude, and duration. We show how the parameters of our signal model may allow us to place constraints on the proto-neutron star mass and radius, the turbulent kinetic energy onto the proto-neutron star, and the time of shock revival.

\end{abstract}

\maketitle

\section{Introduction}
\label{sec:intro}  

Advanced gravitational-wave detectors have finished their first three observing runs detecting $\sim\,90$ signals from compact binary systems \cite{2019PhRvX...9c1040A, 2021PhRvX..11b1053A, 2021arXiv211103606T}. The Advanced LIGO \cite{2015CQGra..32g4001L}, Advanced Virgo \cite{2015CQGra..32b4001A} and KAGRA \cite{2020arXiv200505574A} gravitational-wave detectors are now undergoing upgrades that will further increase the detectors sensitivities. As the detectors become more sensitive, they will start to make observations of other sources of gravitational waves. One of those potential sources is a nearby core-collapse supernova (CCSN). They are extremely powerful explosions with energies of typically $10^{51}$\,erg, and may result in the first multi-messenger detection of electromagnetic radiation, gravitational waves and neutrinos. While supernova neutrinos have already been observed in the case of SN~1987A, no CCSN signals were detected during the first few advanced gravitational-wave detector observing runs \cite{2020PhRvD.101h4002A}. 

The gravitational-wave emission produced during a CCSN explosion is predicted through computationally expensive hydrodynamical simulations (See \cite{2020arXiv201004356A} for a recent review). In recent years, a significant number of CCSN gravitational-wave predictions from 3D hydrodynamical simulations have become available \cite{2020PhRvD.102b3027M, 2020MNRAS.494.4665P, 2019MNRAS.486.2238A, 2019MNRAS.487.1178P, 2019ApJ...876L...9R, 2018ApJ...865...81O, 2018ApJ...857...13P}. However, using the gravitational waveforms from CCSN simulations for gravitational-wave parameter estimation is difficult for a number of reasons. Simulating the waveforms in 3D is too computationally expensive to fully cover the CCSN parameter space, and there are no analytical models that can fully describe a CCSN signal for any given set of parameters. The time series waveforms have a stochastic phase component that make it impossible to produce time series templates for the full CCSN signals. 

In previous CCSN parameter estimation studies, the time series gravitational-wave signal around the core-bounce time of the CCSN have been used to predict the rotation rate and equation of state (EoS) of the progenitor \cite{2009PhRvD..80j2004R, 2021PhRvD.103b4025E, 2014PhRvD..90d4001A, 2017PhRvD..95f3019R, 2014InvPr..30k4008E, 2014PhRvD..90l4026E, 2008ApJ...678.1142S}. However, at later times after the core-bounce, the time series contains stochastic elements that make it extremely difficult to relate the parameters of the time series signal back to the properties of the source. This issue may be avoided as common features of CCSN explosions, that are related to the source properties, can be determined from the spectrogram of the gravitational-wave signal. These features include high-frequency modes that are related to the mass and radius of the proto-neutron star (PNS), and before shock revival there can be lower frequency modes due to the standing accretion shock instability (SASI) \cite{0004-637X-584-2-971, 2006ApJ...642..401B, 2007ApJ...654.1006F}. 

Current gravitational-wave search and waveform reconstruction techniques for CCSN signals do not include knowledge of their known signal features, as they use signal models that make minimal assumptions about the features of the signal \cite{2021arXiv210406462S, 2015CQGra..32m5012C}. However, work has begun to transform the 
known features of the signal spectrograms
into phenomenological models that can be used for gravitational-wave searches and parameter estimation. Using 2D simulations, Morozova et al. \cite{2018ApJ...861...10M}, study the gravitational-wave g/f modes using a range of masses and equations of state (EoS) and find that they can represent the trajectory of the mode frequency during the first 1.5\,s of the gravitational-wave emission with a simple quadratic equation. This model fits well whilst the progenitor core is shrinking, but eventually the mode frequency will level out when the PNS will cool down and deleptonize. 
Similarly, Warren et al. \cite{2019arXiv191203328W} use a slightly different phenomenological model which they find fits better at later times than the model of Morozova et al. when the gravitational-wave emission stops rising and begins to level out. 
Torres-Forn\'e et al. \cite{2019PhRvL.123e1102T} and Sotani et al. \cite{2021arXiv211003131S} use CCSN waveforms from 2D simulations to produce universal relations that relate the g- p- and f- modes to properties of the source, such as the surface gravity of the PNS or the mean density in the region enclosed by the shock. 
The authors state that their model fits could potentially be used in gravitational-wave data analysis studies, however they do not attempt this in their publications. Bizouard et al. \cite{2021PhRvD.103f3006B}, use the universal relations from Torres-Forn\'e et al. \cite{2019PhRvL.123e1102T}, and a set of 1D CCSN simulations with masses between $11\,M_{\odot}$ and $40\,M_{\odot}$, to infer the PNS properties of 2D CCSN signals in current or next generation gravitational-wave detectors. They use a maximum likelihood approach to fit a polynomial to detect gravitational-wave signals in LIGO or future gravitational-wave detectors.

Srivastava et al. \cite{2019PhRvD.100d3026S} model both the lower frequency component of the CCSN gravitational-wave emission and the high-frequency modes using a different approach. They use a combination of five sine Gaussians with different peak frequencies to create a phenomenological model. 
They use their phenomenological model to optimise future gravitational-wave detector designs for the detection of a CCSN. However, they also do not use their model to attempt CCSN parameter estimation. 

A number of previous studies have modelled the gravitational-wave signal by applying principal component analysis to a selection of time series waveforms from numerical simulations \cite{2009CQGra..26j5005H, 2009PhRvD..80j2004R, 2012PhRvD..86d4023L, 2016PhRvD..94l3012P, 2014PhRvD..90l4026E}. In Roma et al. \cite{2019PhRvD..99f3018R}, they use spectrograms of a linear combination of the first few principal components as their signal model and a non-central chi-squared likelihood function, to remove the issues caused by the stochastic components of the time series, and allow them to carry out their analysis in the spectrogram domain. They then use their models to perform Bayesian model selection to determine if certain modes are present in a gravitational-wave detection. However, their principal component signal model cannot be related back to the astrophysical parameters of the original waveforms used to create the principal components. Coughlin et al. \cite{2014CQGra..31p5012C} use a spectrogram as a signal model for generic sine Gaussian burst parameter estimation.

Astone et al. \cite{2018PhRvD..98l2002A} produce a phenomenological model to describe the high-frequency CCSN signal mode, which they use to train a machine learning algorithm for the detection of CCSN gravitational-wave signals, however they do not perform waveform reconstruction or parameter estimation with their model. In a follow up study, Lopez et al. \cite{2021PhRvD.103f3011L} show that machine learning algorithms trained with this phenomenological model can detect CCSN waveforms from hydrodynamical simulations. 

In this paper, we expand on the previous work by adding a phenomenological model for the high-frequency CCSN gravitational-wave signal mode to the Bayesian parameter estimation code Bilby \cite{2019ApJS..241...27A}. We use an asymmetric chirplet model to represent the dominant  gravitational-wave emission mode. We use the non-central chi-squared likelihood function from \cite{2019PhRvD..99f3018R} to allow us for the first time to perform our CCSN parameter estimation analysis in the spectrogram domain instead of using the stochastic time series waveforms. We show the image reconstruction of the gravitational-wave modes, and we determine how well we can estimate the parameters of our signals such as their duration and peak frequency. We then show how these parameters measured with our phenomenological model can inform us of the astrophysical properties of the source such as the mass and radius of the PNS, the time of shock revival and the turbulent kinetic energy onto the PNS. 

This paper is structured as follows: In Section \ref{sec:waveforms}, we give a brief description of the simulated CCSN gravitational-wave signals that we use to test our method. In Section \ref{sec:pe}, we give the details of our new asymmetric chirplet  signal model and the non-central chi-squared likelihood that we use in our analysis. In Section \ref{sec:results}, we give our parameter estimation results and explain how these results can be converted into posteriors on astrophysical properties of the source. A discussion and conclusion are given in Section \ref{sec:conclusion}.

\section{Supernova gravitational-wave signals}
\label{sec:waveforms}

\begin{figure*}
\includegraphics[width=\textwidth]{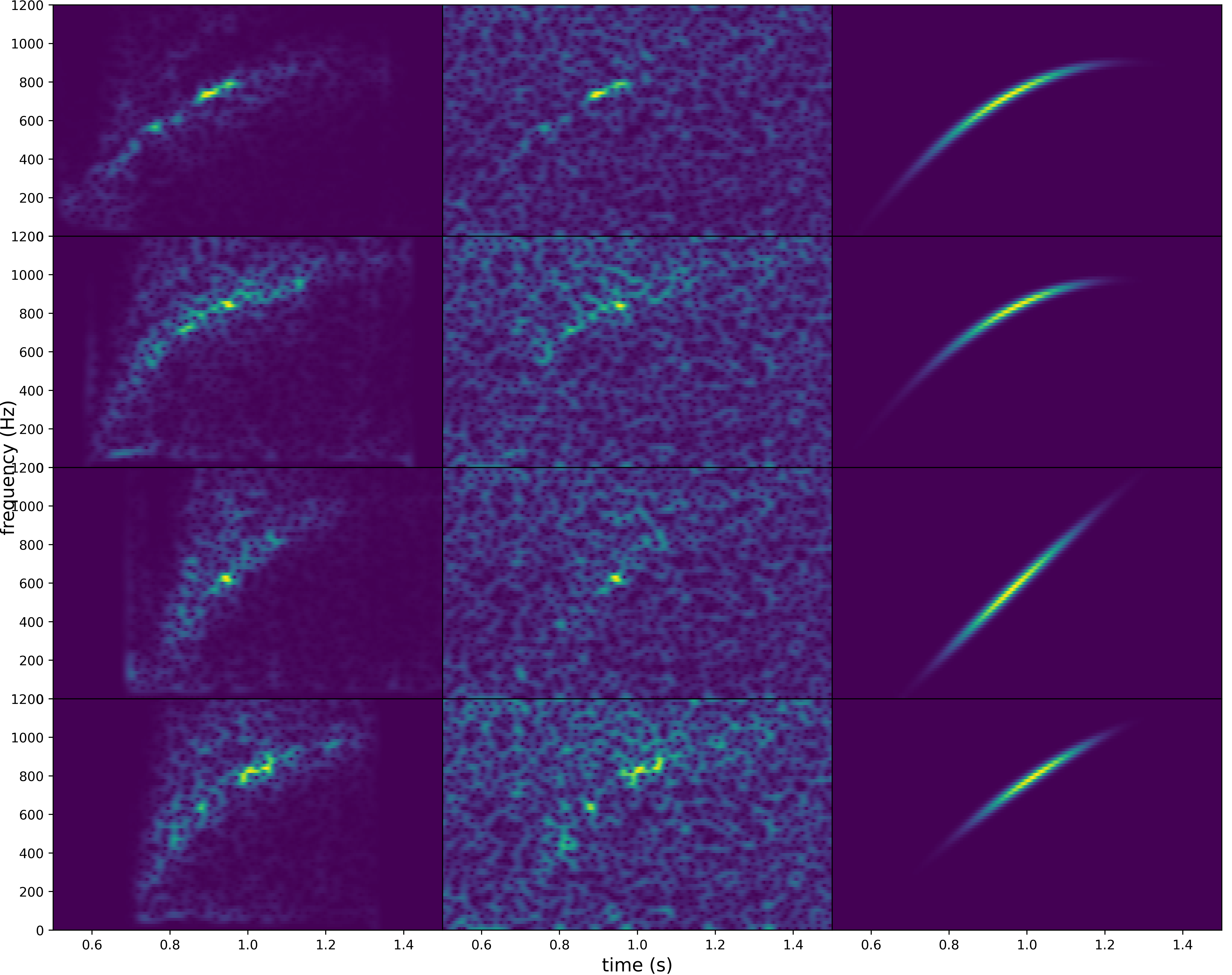}
\caption{From top to bottom are models m39, s18, y20, he3.5. From left to right the columns show the waveform from simulations, the waveform in Advanced LIGO design sensitivity noise \cite{2015CQGra..32g4001L}, and the reconstructed waveform. The signal has a signal to noise ratio of 25. Our signal model captures the shape of the dominant signal mode but not the stochastic fluctuations in amplitude.  }
\label{fig:recon}
\end{figure*}

To test our method, we use four different CCSN signals from our recent simulations with the neutrino hydrodynamics code CoCoNuT-FMT \cite{2002A&A...393..523D, 2010ApJS..189..104M}. We choose a range of different progenitor masses, and include one model with rotation, to provide a good representation of the full CCSN parameter space. Spectrograms of all of the simulated gravitational-wave signals are shown in Figure \ref{fig:recon}, and further details of each model are provided in the sub-sections below.

\subsection{Model m39}

Model m39 is a rapidly rotating Wolf-Rayet star with an initial helium star mass of $39\,M_{\odot}$ \cite{aguilera_18,2020MNRAS.494.4665P}. It has an initial surface rotation velocity of $600\,\mathrm{kms}^{-1}$ and has a metallicity of $1/50\,Z_{\odot}$. Due to the rapid rotation, this model achieves shock revival shortly after core-bounce, and reaches a high explosion energy of $11\times10^{50}$\,erg by the end of the simulation, which results in high amplitude gravitational-wave emission. The simulation ends 0.98\,s after the core-bounce. The amplitude of the gravitational-wave emission peaks at a frequency of 733\,Hz.

\subsection{Model s18}

Model s18 is a solar-metallicity star with a zero-age main sequence (ZAMS) mass of $18\,M_{\odot}$ \cite{2019MNRAS.487.1178P}. The simulation was stopped 0.89\,s after core bounce. The amplitude of the gravitational-wave signal is highest at a peak frequency 894\,Hz. The shock is revived at $\sim0.3$\,s after the core bounce and by the end of the simulation has reached an explosion energy of $3.3\times10^{50}$\,erg.

\subsection{Model y20}

Model y20 is a $20\,M_{\odot}$, non-rotating, solar-metallicity helium star \cite{yoon_17,2020MNRAS.494.4665P}. The shock is revived $\sim 200$\,ms after core-bounce and reaches an explosion energy of $5.9\times10^{50}$\,erg. The amplitude of the gravitational-wave signal is highest at a peak frequency of 606\,Hz. The simulation was performed for a  duration of 1.2\,s after the core-bounce time, however the gravitational-wave amplitude is only high for the first 0.6\,s. 

\subsection{Model he3.5}

Model he3.5 is an ultra-stripped star with a helium core mass of $3.5\,M_{\odot}$ \cite{tauris_15,2019MNRAS.487.1178P}. The simulation is stopped 0.7\,s after core-bounce. The amplitude of the gravitational-wave signal is highest at a peak frequency of 824\,Hz. The shock is revived $\sim 400$\,ms after the core-bounce. However, due to the low mass of this model, it only reaches an explosion energy of $0.99\times10^{50}$\,erg, and has lower gravitational-wave amplitudes than the other models.

\section{Parameter estimation method}
\label{sec:pe}

Before we start the analysis we do some pre-processing to the CCSN waveforms. First we resample the waveforms to a frequency of 4096\,Hz, apply a high pass filter to remove frequencies below 30\,Hz, and zero pad them so they are 2\,s long and scale them to the relevant distance. We use simulated Gaussian design sensitivity Advanced LIGO detector noise, for both LIGO Livingston and LIGO Hanford.

To perform our Bayesian parameter estimation we use the Python Bayesian inference code Bilby \cite{2019ApJS..241...27A, 2020MNRAS.499.3295R}, which is used for the parameter estimation of compact binary coalesences in advanced gravitational-wave detectors. Bilby can produce Bayes factors that can be used for model selection, and can also produce posterior distributions on signal model parameters $\theta$. The posteriors are calculated using Bayes' theorem where for data $d$ and model $\mathcal{H}$,
\begin{equation}
p(\theta|d,\mathcal{H}) = \frac{ \mathcal{L}(d|\theta,\mathcal{H}) \pi (\theta|\mathcal{H}) }{ \mathcal{Z}(d|\mathcal{H}) }   ,
\end{equation}
where $\mathcal{L}(d|\theta,\mathcal{H})$ is the likelihood, $\pi (\theta|\mathcal{H})$ is the prior and $\mathcal{Z}(d|\mathcal{H})$ is the evidence. The prior includes our knowledge of the parameters before the start of the analysis. The evidence is given by, 
\begin{equation}
\mathcal{Z}(d|\mathcal{H}) = \int p(d|\theta,\mathcal{H}) \pi (\theta|\mathcal{H}) d\theta   .
\end{equation}
It is a normalisation constant for parameter estimation and is important in model selection. The standard Gaussian likelihood function used in gravitational-wave parameter estimation is described in \cite{PhysRevD.46.5236}, however we add a new likelihood function and signal model to Bilby in order to perform our analysis in the spectrogram domain. 

The signal model is an asymmetric chirplet model \cite{DEMIRLI2014907} defined in the time domain as,  
\begin{equation}
h_{+} = A \times \exp\left(-\frac{bdt^2}{\tau^2}\right) \times \cos(2\pi f dt + 2 \pi \dot{f} dt + c dt^3)
\end{equation}
\begin{equation}
h_{\times} =  A \times \exp\left(-\frac{b dt^2}{\tau^2}\right) \times \sin(2\pi f dt + 2 \pi \dot{f} dt + c dt^3)   
\end{equation}
where A is the amplitude, $f$ is the peak frequency, which is the frequency where the amplitude of the signal is highest, $\dot{f}$ is the rate of change of frequency, $dt$ is the array of time steps minus the time where the center of the signal occurs, $c$ is a parameter that controls how curved the chirplet is. This extra $c$ parameter allows the end of the signal to begin to level off instead of continuing to rise in frequency with time, as this is what we see in waveform simulations. The parameter $\tau$ is defined at $\tau = Q / 2 \pi f$, where Q is the number of cycles. The parameter $b$ is given by, 
\begin{equation}
b = 1 - a \tanh(dt)    ,
\end{equation}
where $a$ is the asymmetry parameter that allows one side of the chirplet to be larger than the other as,
\begin{equation}
a = \frac{\alpha_L - \alpha_R}{\alpha_L + \alpha_R}    ,
\end{equation}
where $\alpha_L$ is decay rate for the left half of the chirplet and $\alpha_R$ is the decay rate for the right half of the chirplet. 
The root sum squared amplitude $h_\mathrm{rss}$ is given by,
\begin{equation}
h_\mathrm{rss} = \sqrt{ \sum_{\tau} ( |h_+|^{2} + |h_{\times}|^{2} ) dt} ,
\end{equation}
where $\tau$ is the duration, and $dt$ is one over the sample rate. 
Circularly polarised gravitational-wave signals are not very realistic for CCSNe, however they are standard in most gravitational-wave burst algorithms \cite{2020arXiv200612604D, 2015CQGra..32m5012C, 2015arXiv151105955L}, and do not change the results in our case due to performing our analysis using spectrograms. 
For all the results in this study, we use the sky position of the Galactic center, and we assume that the sky position and the distance of the source are already known either from electromagnetic observations or neutrinos. 
We then convert the time domain asymmetric chirplet into a spectrogram with segment length 128, and the number of overlap points equal to 65. 

To avoid the issue of the stochastic phase in CCSN waveforms, and to allow us to use spectrograms as our signal model, we use a non-central chi-squared likelihood function given by, 
\begin{multline}
\log L_{S} = N \log \left(\frac{1}{2}\right) + \\ \sum_{n=1}^N \left( \frac{-(d^2_n+h^2_n)}{S(f)} + \log\left(I_o \left[\frac{2d_nh_n}{S(f)}\right] \right)  \right)    ,
\end{multline}
where $d$ is the spectrogram of the detector data, $I_o$ is the modified Bessel function, $h$ is the spectrogram of the signal model, $S(f)$ is the noise power spectral density, and $N$ is the total number of frequency bins with corresponding index $n$. The noise-only likelihood function is given by, 
\begin{equation}
\log L_{N} = N \log \left(\frac{1}{2}\right) -\sum_{n=1}^N \left( \frac{d^2_n}{S(f)} \right) .   
\end{equation}
This is the same likelihood function that was used in \cite{2019PhRvD..99f3018R}.  

We use uniform priors on all the parameters in our model. We assume the peak frequency is between 500\,Hz and 1000\,Hz, as this is consistent with the value observed in previous 3D waveform simulations. We assume a maximum signal duration of 1.1\,s, and a maximum $h_\mathrm{rss}$ of $1\times10^{21}$. The asymmetry parameter prior is uniform between -1 and 1. Although the core-bounce time of the signal may be well known if a neutrino counterpart is detected, the central time of our model is defined as the time where the amplitude of the signal is highest, which is likely to occur shortly after the shock revival time. 
As this time is more uncertain than the core-bounce time, we allow our model to vary the central time by half a second.

\section{Results}
\label{sec:results}

\subsection{Signal Detection}

\begin{figure}
\includegraphics[width=\columnwidth]{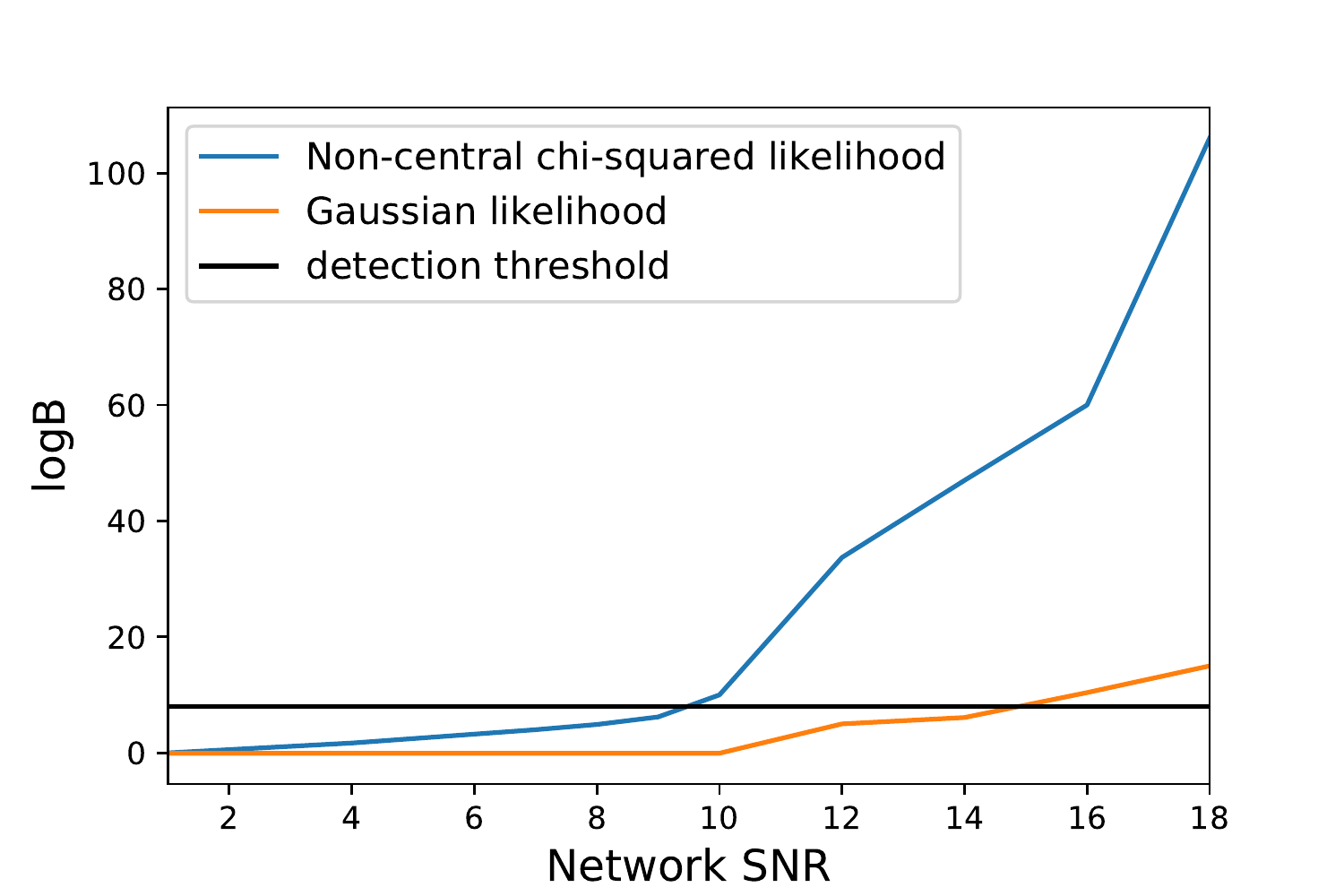}
\caption{The minimum Network SNR needed to detect the m39 model using the non-central chi-squared likelihood and spectrogram signal model, and the standard Gaussian likelihood with the time domain signal model. Performing the analysis in the spectrogram domain significantly decreases the minimum SNR needed for detection.}
\label{fig:detect}
\end{figure}

To be able to determine the parameters of a CCSN gravitational-wave signal, first the parameter estimation code needs to be able to find the signal in the data. It is common in gravitational-wave astronomy to consider a log Bayes factor of 8 as the detection threshold \cite{2015PhRvD..91d2003V}. In Figure \ref{fig:detect}, we show the minimum network SNR needed to be able to detect the m39 signal at this threshold, where the network SNR is given by,
\begin{equation}
\mathrm{SNR}_{net} = \sqrt{\mathrm{SNR}_{L1}^{2} + \mathrm{SNR}_{H1}^{2}},
\end{equation}
where $\mathrm{SNR}_{L1}$ is the SNR in the LIGO Livingston detector, and $\mathrm{SNR}_{H1}$ is the SNR in the LIGO Hanford detector.
We use our asymmetric chirplet signal model and use both the non-central chi-squared likelihood and the standard Gaussian likelihood function in Bilby to see if the new likelihood function improves the detection threshold for CCSN signals. We find using the Gaussian likelihood, and a time domain asymmetric chirplet, that we require a minimum detector network SNR of 15 to be able to detect the signal. Using the non-central chi-squared likelihood function reduces the required network SNR to 10. This shows that using the non-central chi-squared likelihood function, which removes the stochastic phase elements of the signals, will allow us to perform Bayesian parameter estimation for sources at greater distances. Similar results are found for all the models that we consider in this study.

\subsection{ Mode Reconstruction }

\begin{figure}
\includegraphics[width=\columnwidth]{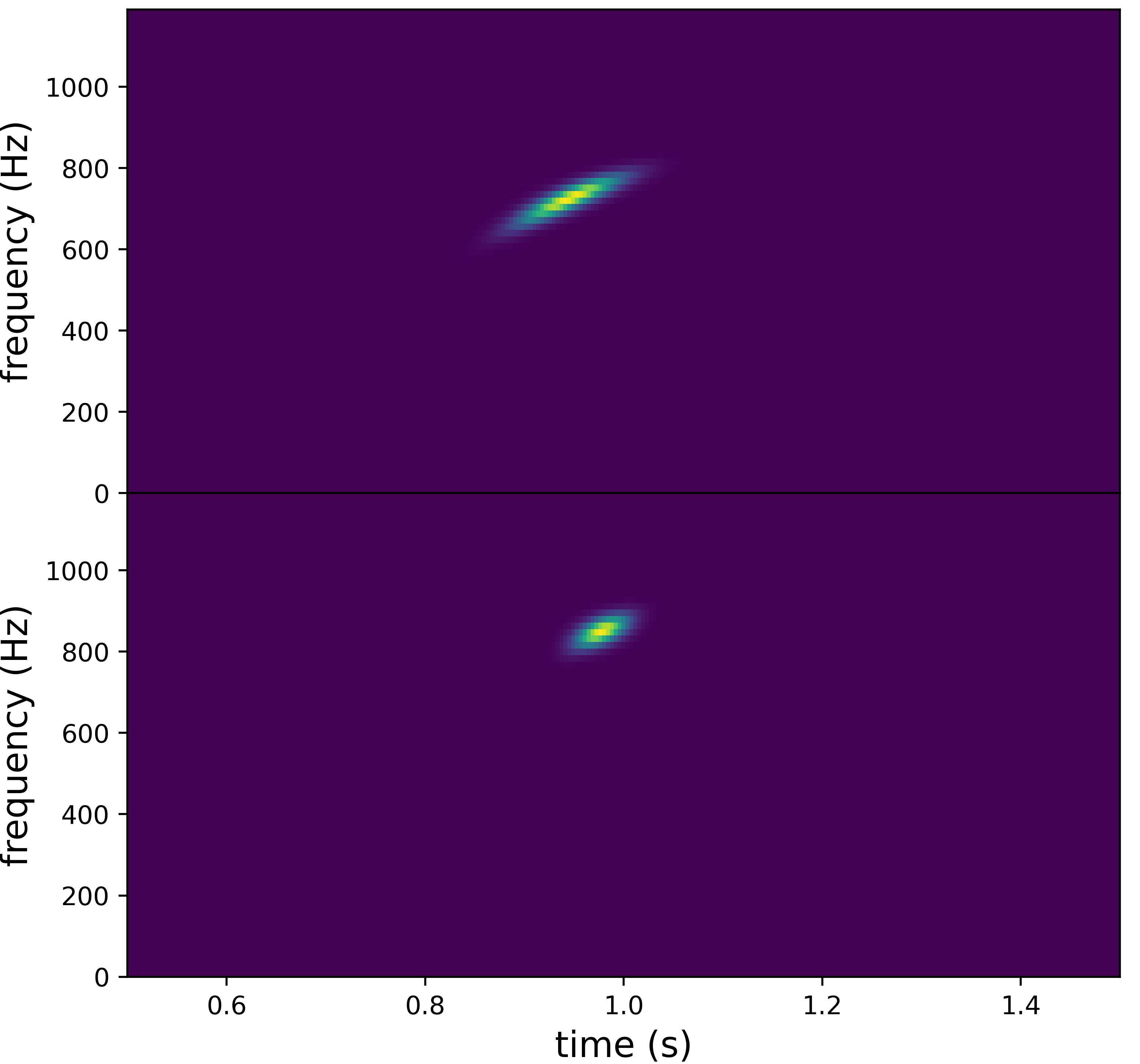}
\caption{The reconstructed waveforms for m39 (top) and s18 (bottom) using the Gaussian likelihood function and time domain signal model. The network signal to noise ratio is 25. 
}
\label{fig:oldlogl}
\end{figure}

In Figure \ref{fig:recon}, we show an example reconstructed waveform for all four signal models at a network SNR of 25. Our model is able to produce a good representation of the gravitational-wave signal mode for all of the models, but it cannot reproduce the individual fluctuations in amplitude along the mode. This may be possible with other gravitational-wave burst signal reconstruction tools \cite{2015CQGra..32m5012C, 2021arXiv210406462S}, however the mode parameters reconstructed by our model are enough to rapidly infer astrophysical information about the source. The stochastic fluctuations in signal amplitude do not inform us of interesting details about the source astrophysics. 

In Figure \ref{fig:oldlogl}, we show the reconstructions of the m39 and s18 models at SNR 25 when using the Gaussian likelihood and time domain asymmetric chirplet model. Even for loud signals, it is not possible to capture the entire mode when using the time series signal model and Gaussian likelihood. Only the part of the mode where the signal amplitude is highest is captured by the time series signal model.

\subsection{ Source Properties }

\begin{figure*}
\includegraphics[width=\columnwidth]{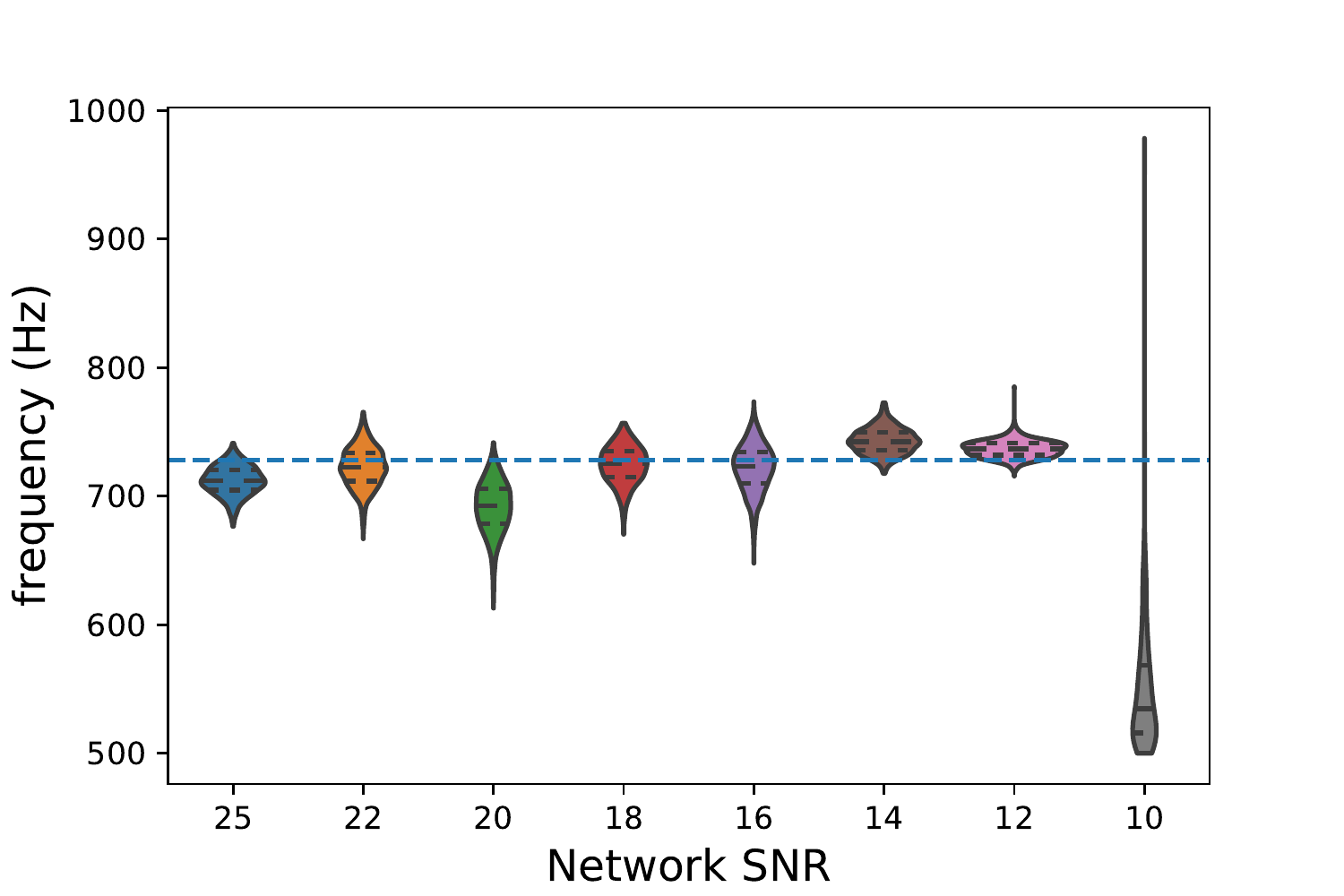}
\includegraphics[width=\columnwidth]{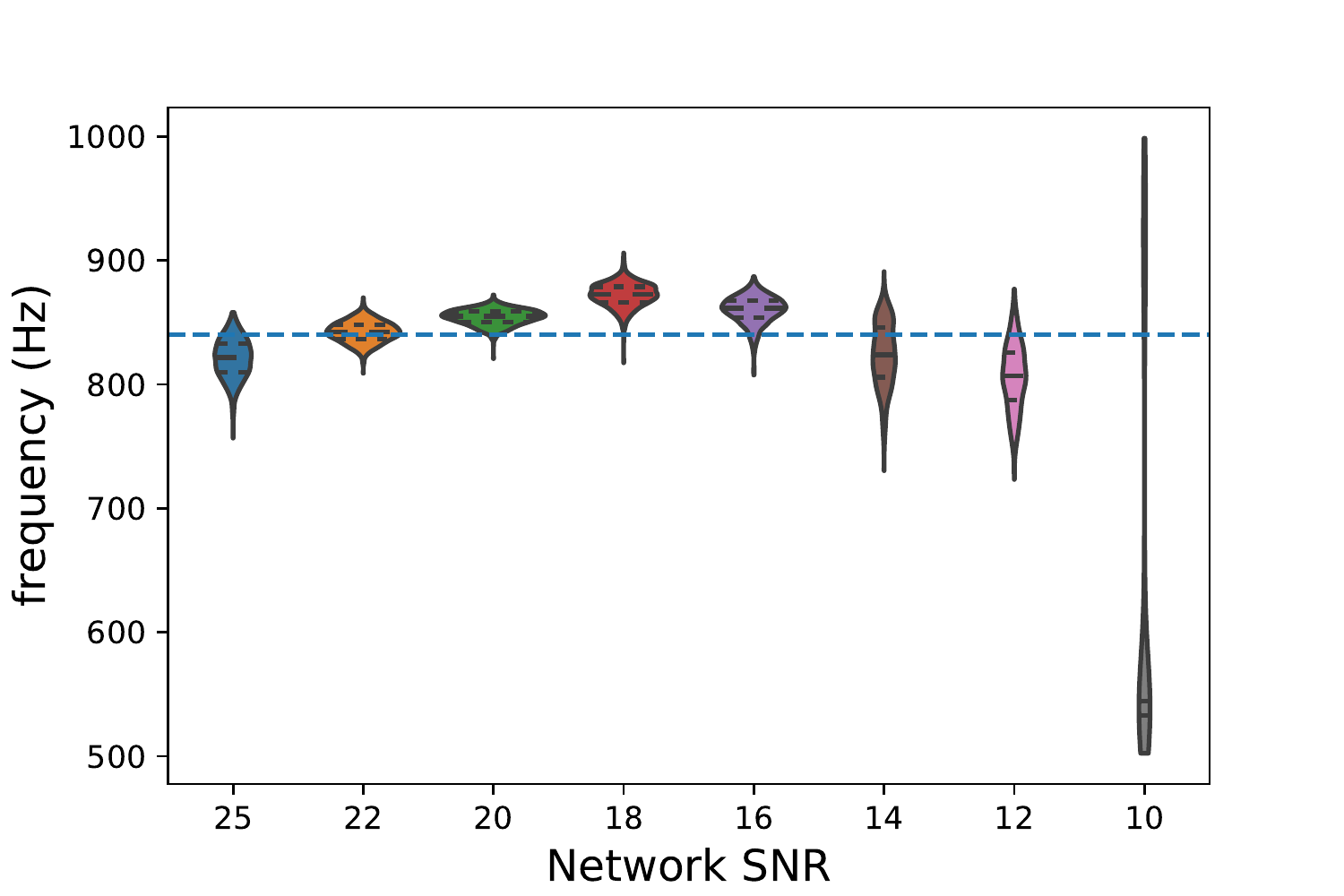}
\includegraphics[width=\columnwidth]{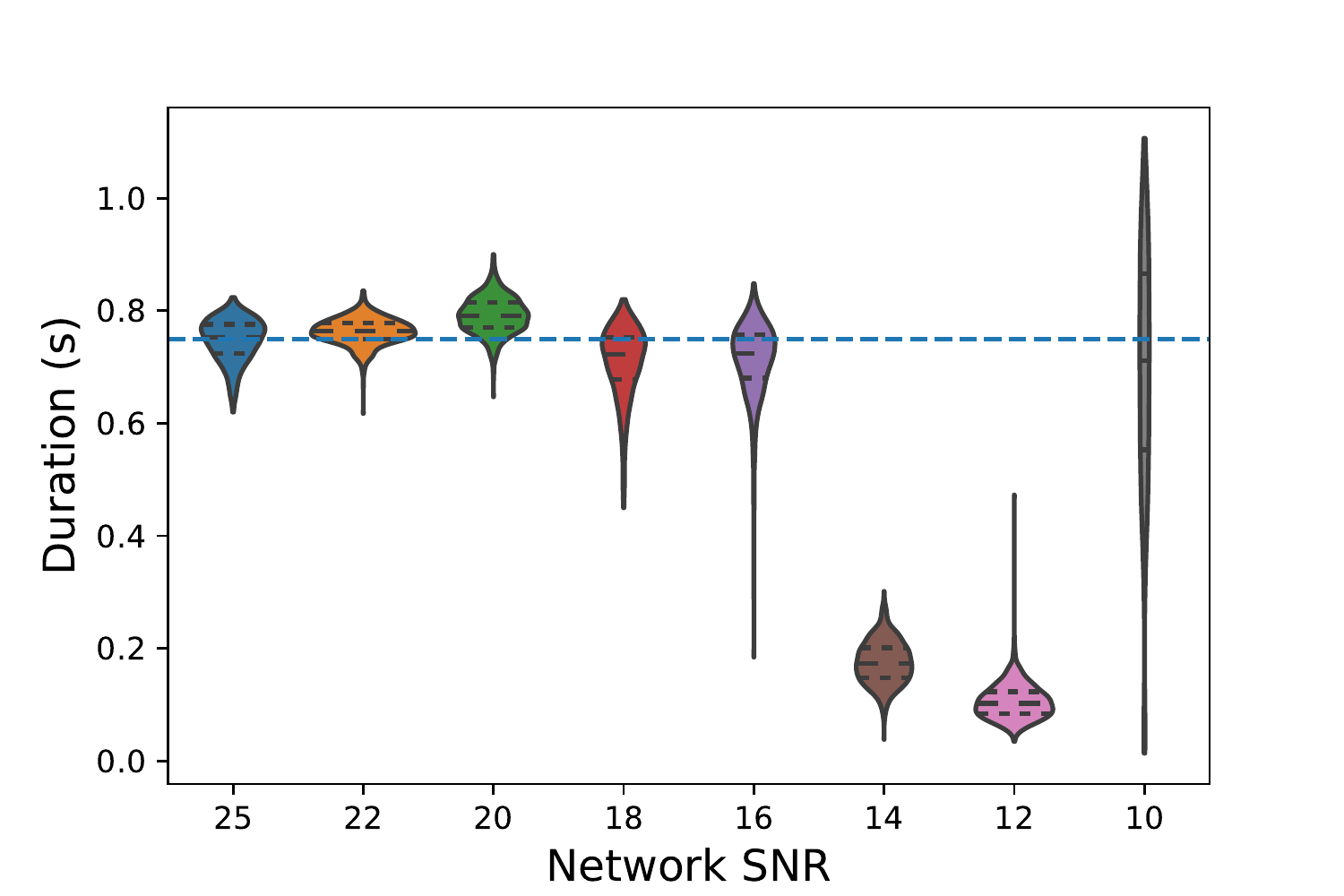}
\includegraphics[width=\columnwidth]{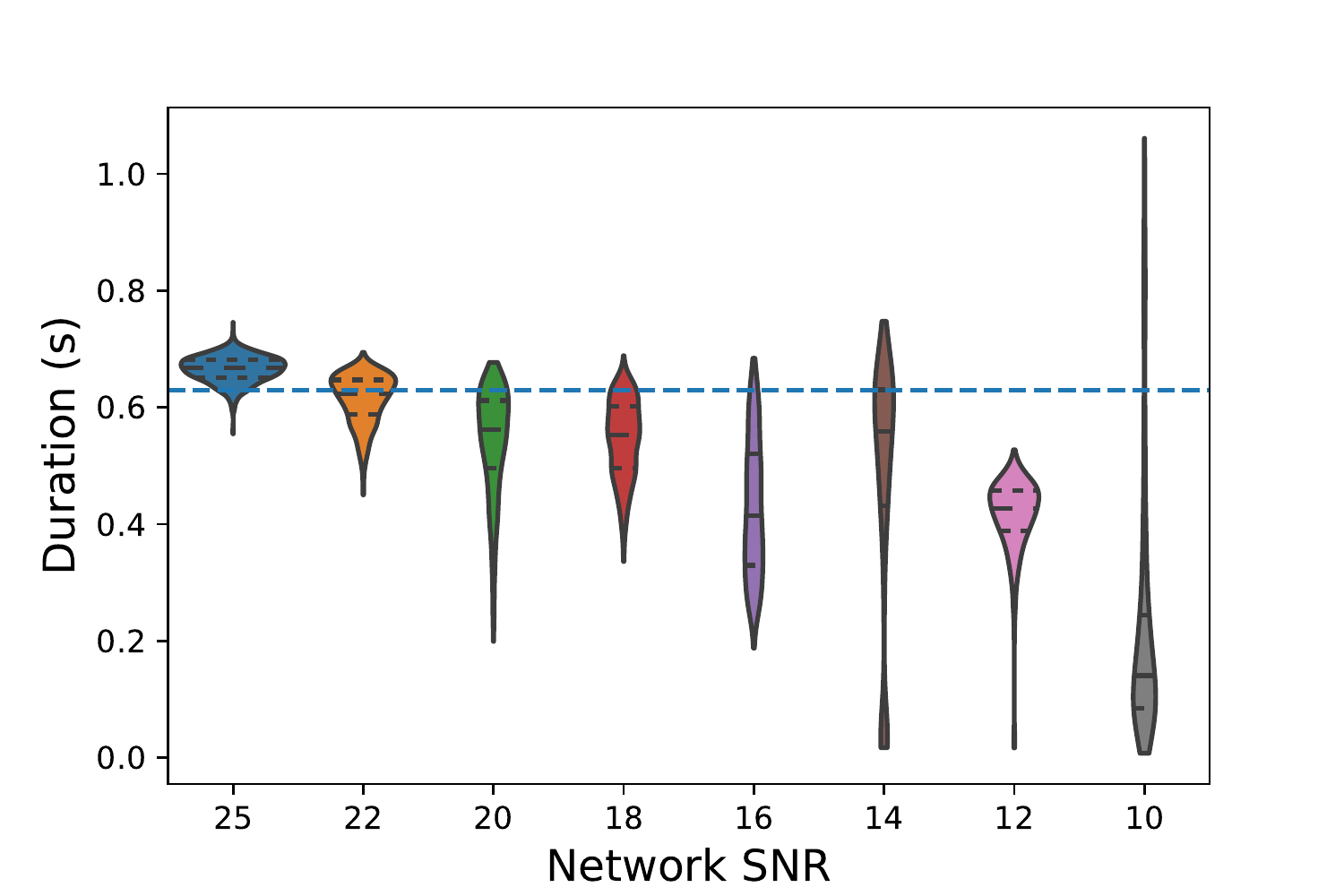}
\includegraphics[width=\columnwidth]{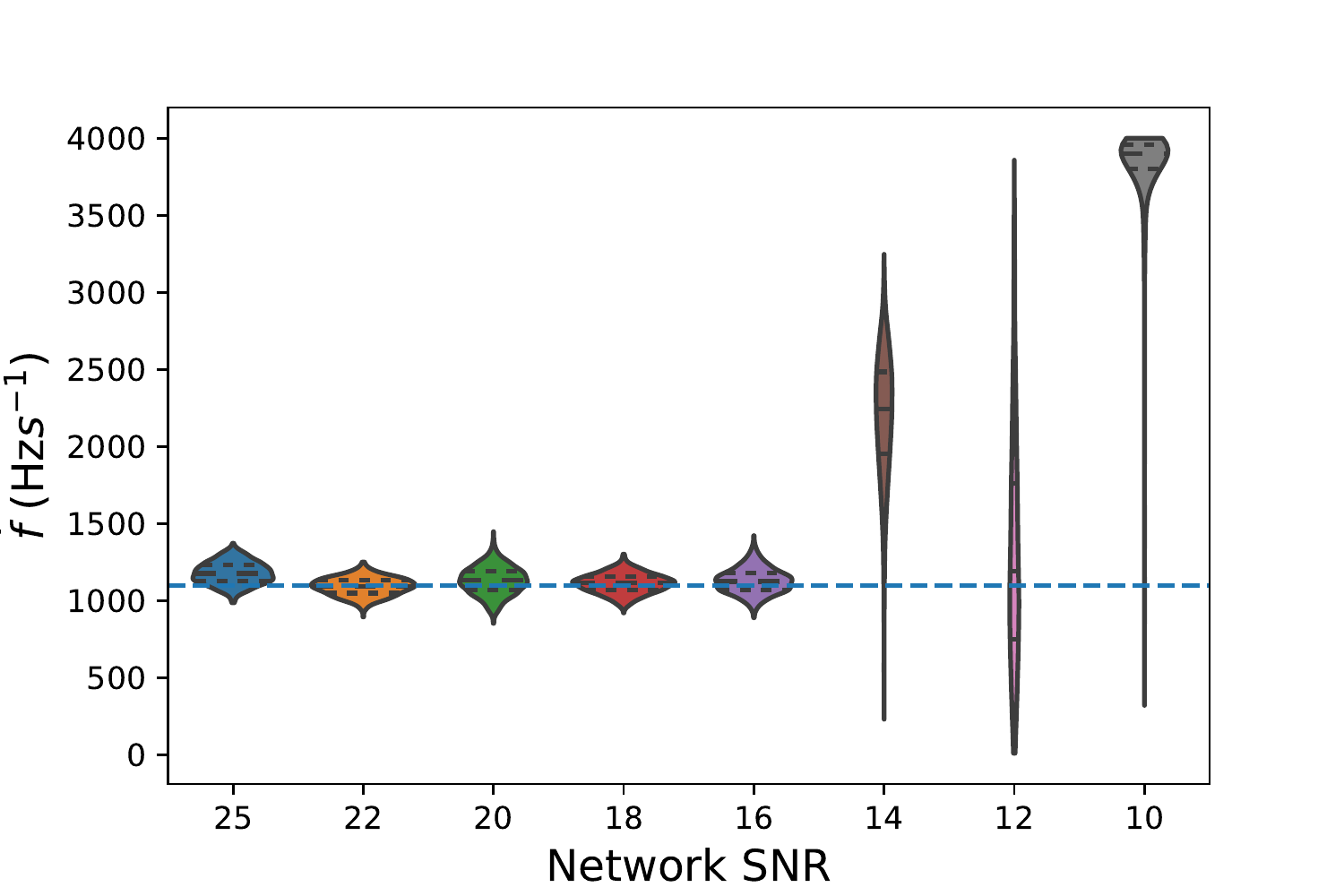}
\includegraphics[width=\columnwidth]{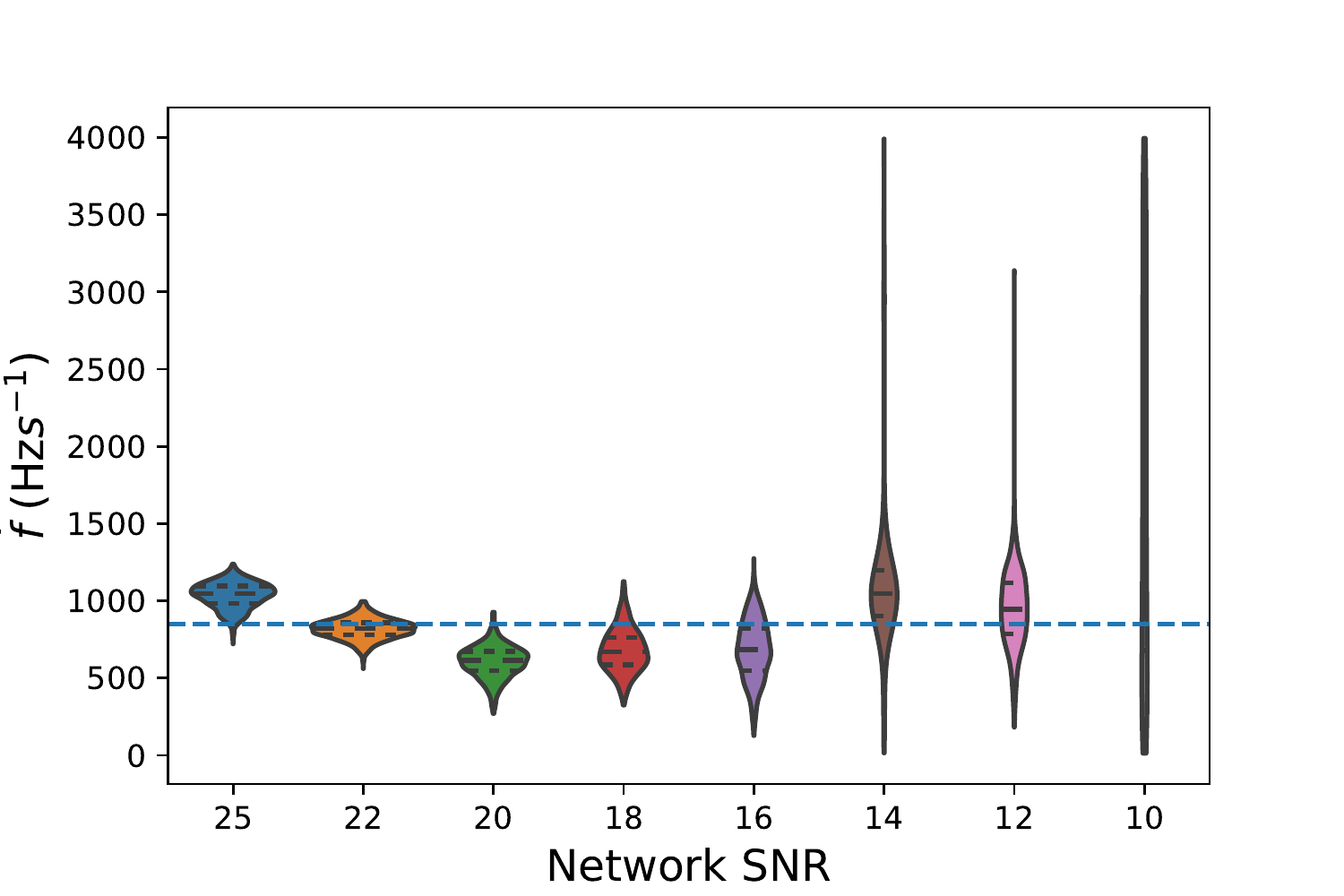}
\caption{The posterior distributions of the peak frequency, duration and rate of change of frequency at different signal to noise ratios. The left column is for model m39, and the right column is for model s18. The dashed lines show the true values. The model parameters are not constrained at a network SNR of 10. At SNR 12 and 14 only the part of the signal with the highest amplitude is reconstructed. SNR 16 is needed to capture the full length of the signal mode.}
\label{fig:params}
\end{figure*}

In Figure \ref{fig:params}, we show some example posterior distributions for the peak frequency, the rate of change of frequency ($\dot{f}$) and the duration. The duration is defined as the time where the amplitude of the gravitational-wave emission is larger than 20\% of the total amplitude. A CCSN could emit gravitational waves for several seconds but the amplitude of the gravitational waves would be too low to be detected. At a network SNR of 10, although the signal can be detected, the error on the parameter measurements is the width of the entire prior used in our analysis, so it would not be possible to make a conclusive statement about the source properties. This is due to the difficulty in distinguishing the signal features from the noise at low SNR. At network SNRs of 12-14, the measured signal duration is short, as the signal is only reconstructed at the times where the waveform has its highest amplitude, which is typically shortly after the shock is revived. This is because the lower amplitude parts of the signal still cannot be distinguished from the noise. This means that the amount of time for which we can predict the values of the PNS mass and radius will be shorter at lower SNR values. 
The peak frequency is well constrained even when the SNR is low, as this occurs when the signal amplitude is highest. 
The $\dot{f}$ parameter is also only well constrained after reaching an SNR of 16. The error in $\dot{f}$ at lower SNR values is due to the smaller duration of the detectable signal.

Measuring the frequency of the mode is important as the mode frequency is related to the mass and radius of the PNS forming during the explosion. Several studies have shown how the gravitational-wave frequency is related to the emission modes visible in the signal \cite{2019PhRvL.123e1102T, 2021arXiv211003131S}. To show how the frequency reconstructed by our analysis can inform us of the properties of the PNS, we use the universal relation for the $^{2}g_{2}$ mode from Torres-Forn\'e et al. \cite{2019PhRvL.123e1102T}. We do not include model m39 as the effects of rotation on the relationship between the gravitational-wave frequency and PNS properties is currently not well understood. 

\begin{figure*}
\includegraphics[width=\columnwidth]{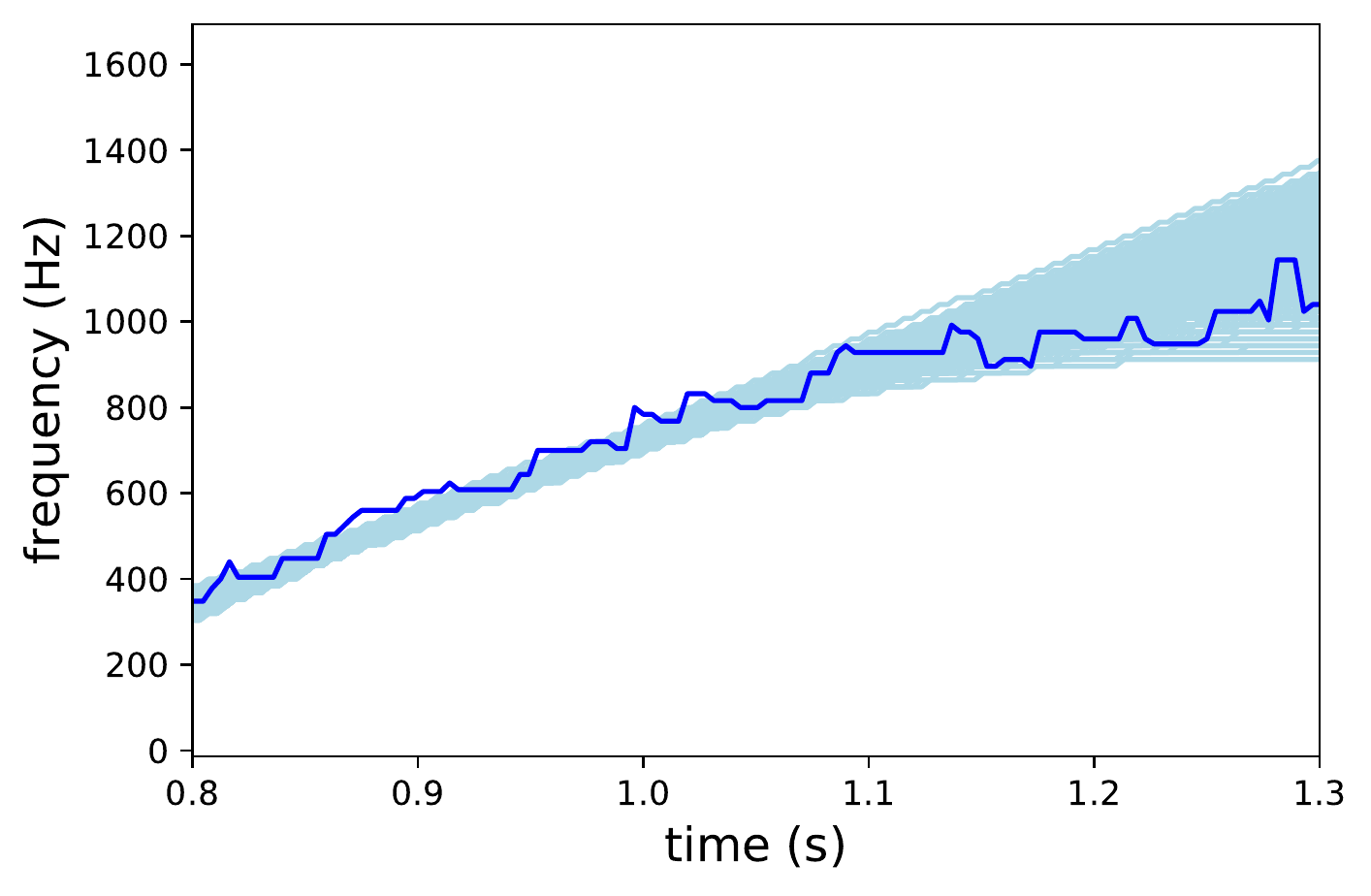}
\includegraphics[width=\columnwidth]{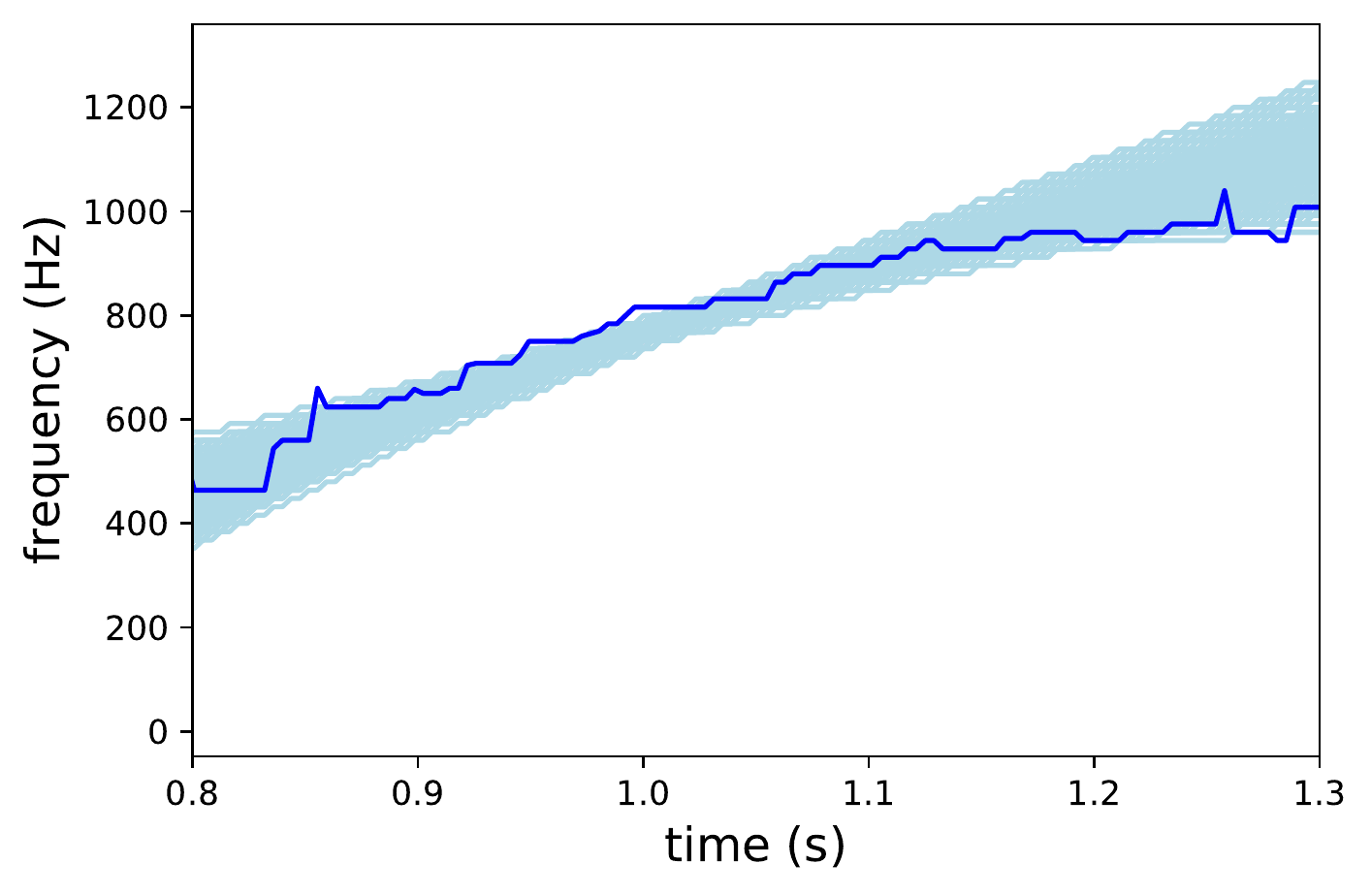}
\includegraphics[width=\columnwidth]{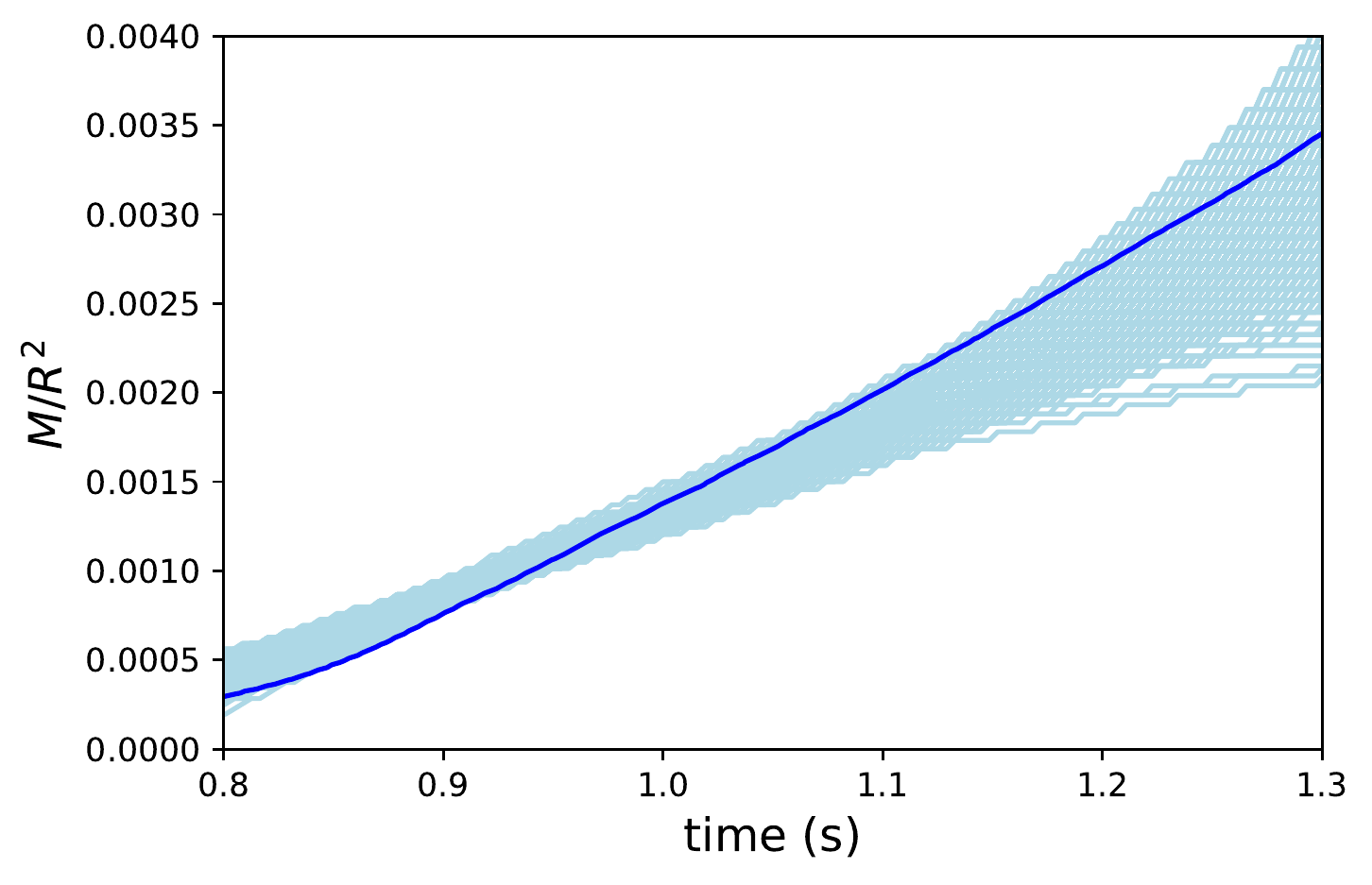}
\includegraphics[width=\columnwidth]{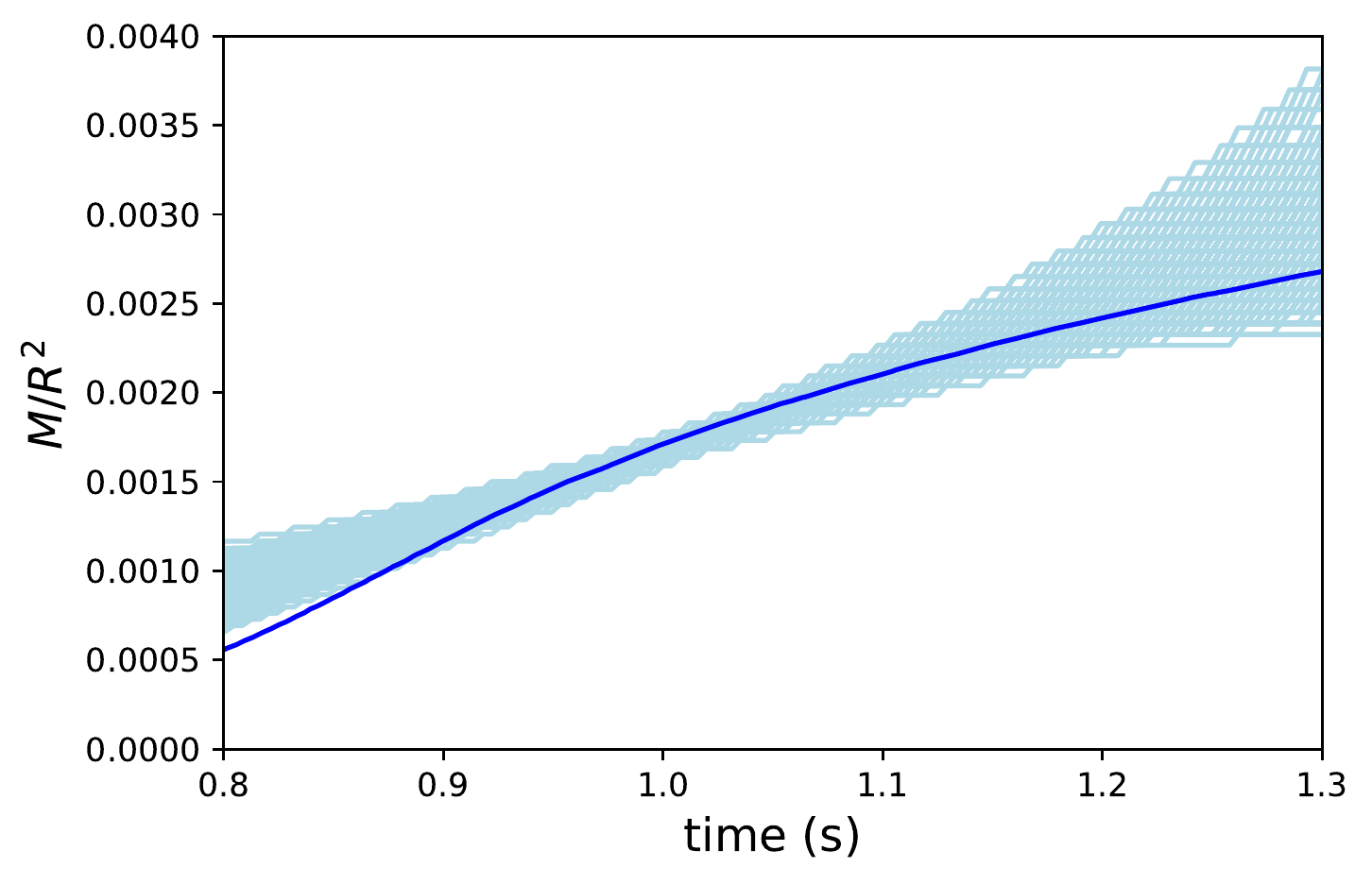}
\caption{Using the frequencies from our reconstructed signal spectrograms, shown in the top row, at an SNR of 25, we estimate the combined proto-neutron star mass and radius $M/R^2$ using the $^{2}g_{2}$ universal relation from Torres-Forn\'e et al. \cite{2019PhRvL.123e1102T}. The light blue shows the values estimated using the posterior samples from Bilby, and the dark blue line is the true value from our waveform simulations. Left is model y20 and right is model he3.5.}
\label{fig:modes}
\end{figure*}

The results are shown in Figure \ref{fig:modes} for models y20 and he3.5. Similar results are also found for model s18. We use a Python root finder to solve the universal relation equation. The equation for the $^{2}g_{2}$ mode frequency $f$ is given by 
\begin{equation}
f = 5.88\times10^{5} x - 86.2\times10^{6}x^{2} + 4.67\times10^{9}x^{3},
\end{equation}
where $x$ is the PNS mass $M$ in solar masses divided by the square of the radius $R$ in km.
To obtain the mode frequency $f$ values, we make spectrograms of an SNR 25 reconstructed signal using all the Bilby posterior samples. The mode $f$ values are then the spectrogram frequency with the highest energy at each time bin. 
We show how well we measure $M/R^2$ for half a second around the time when the gravitational-wave amplitude is highest. The figure shows the true values from the waveform simulations and the light blue lines are made from the Bilby posterior samples. The predicted values are most accurate when the gravitational-wave amplitude is high. The error in the measured value is larger at early times and late times in the signal when the gravitational-wave amplitude is low. As the frequency values of the reconstructed mode are always a good fit for the true values, as shown in the top row of Figure \ref{fig:modes}, it is likely the error in $M/R^2$ is due to the Universal relations being a poor fit to our simulated CCSN signals at early and late times in the signal. At high frequencies, the error on the frequency measurement of the mode is larger, as shown by the larger width of the light blue area in the figure. This is likely due to the much poorer quality of the gravitational-wave noise at higher frequencies. The $^{2}g_{2}$ mode equation also has some error on the numerical parameters in the equation as given in \cite{2019PhRvL.123e1102T}. As our parameter estimation method uses images, there is some limit on how well we can measure the frequency due to the size of our spectrogram frequency bins, which is why in the figure the light blue lines from our analysis are not as smooth as the dark blue line from the CCSN simulations.  

In Bizouard et al. \cite{2021PhRvD.103f3006B}, they also use the same universal relation equation to produce a reconstruction of $M/R^2$ using a single Advanced LIGO detector. However, in their work they have a more consistent level of error for the entire frequency range of the g-mode. This is likely due to the difference in the simulated supernova signals used in their study and they also use a frequency dependent error model of the universal relations. They used 2D waveforms with a more constant gravitational-wave amplitude along the entire duration of the g-mode, where as the 3D waveforms we use in our study have lower gravitational-wave amplitudes at the early and late times in the g-mode.  

\begin{figure}
\includegraphics[width=\columnwidth]{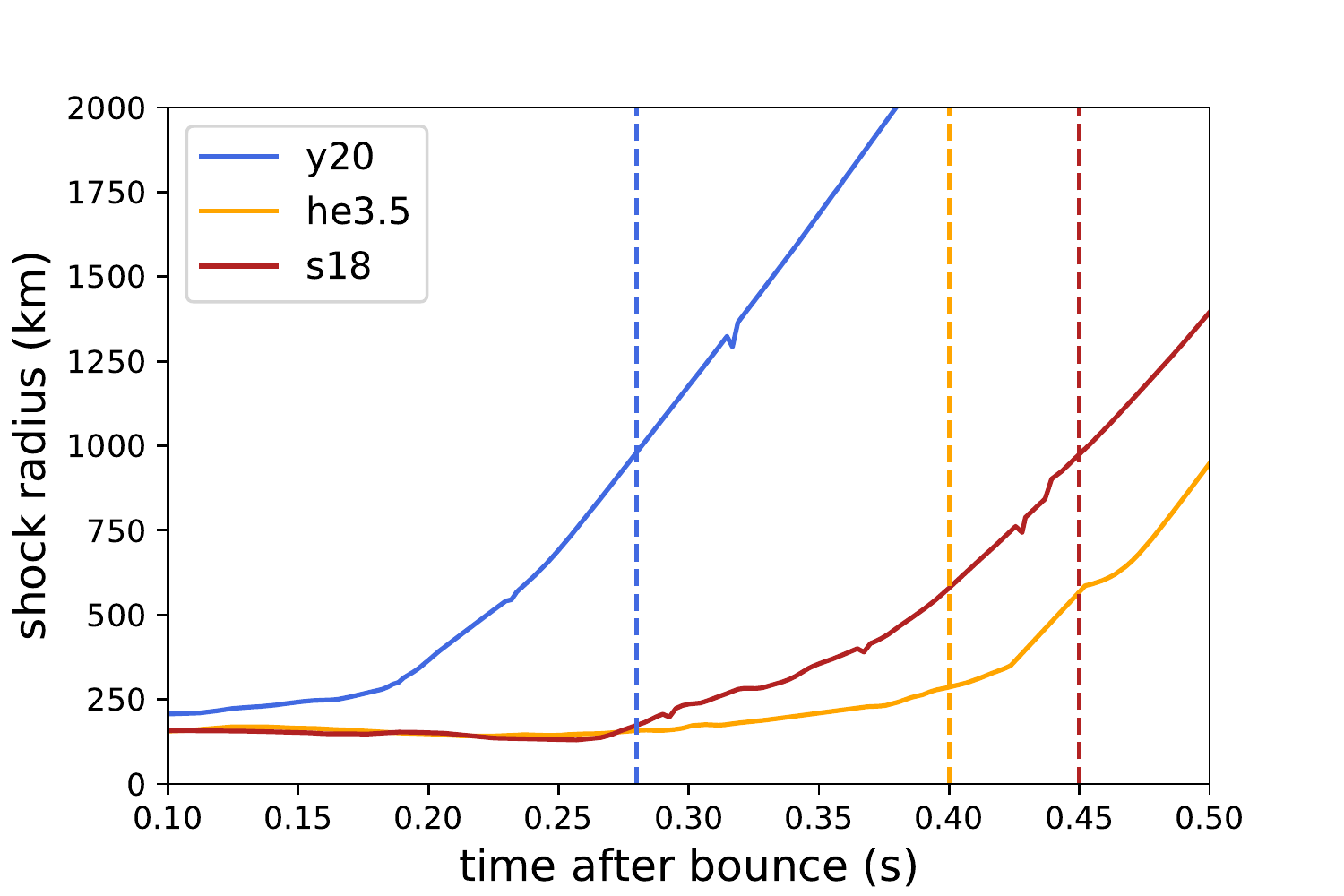}
\caption{The solid lines show the shock radius from the simulations of models y20, he3.5 and s18. The dashed line shows the time of peak frequency measured by our Bilby analysis. Low mass models have their peak frequency around the shock revival time, and higher mass models have their peak frequencies $\sim 100$\,ms after the shock is revived. }
\label{fig:shocktime}
\end{figure}

The results from our parameter estimation analysis may also allow us to make an estimation of the shock revival time. In Figure \ref{fig:shocktime}, we show the shock radius from the CCSN simulations for our models and also the time the peak frequency occurred as measured using Bilby. We find that for low mass models the peak frequency occurs around the shock revival time, and for higher mass models the peak frequency occurs $\sim 100$\,ms after the shock revival time. As it may be possible to measure the CCSN progenitor mass using the neutrino or electromagnetic counterpart signal, we may be able to get an estimation of the shock revival time with an error smaller than $\sim 100$\,ms. We only use the high-frequency mode of the gravitational-wave signal in this study, however the low-frequency SASI mode also contains information that will allow us to further constrain the shock revival time. 

\begin{figure*}
\includegraphics[width=\columnwidth]{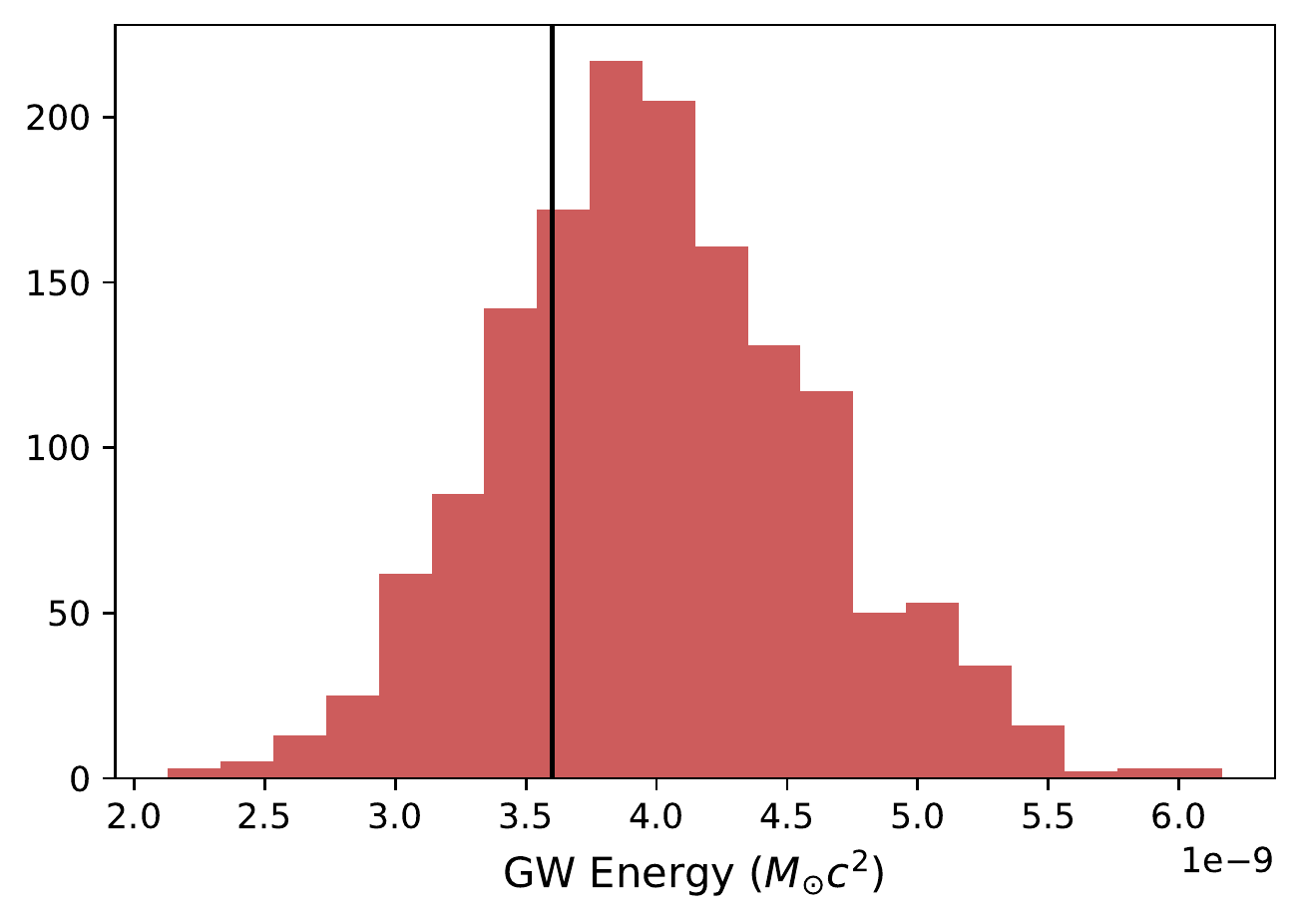}
\includegraphics[width=\columnwidth]{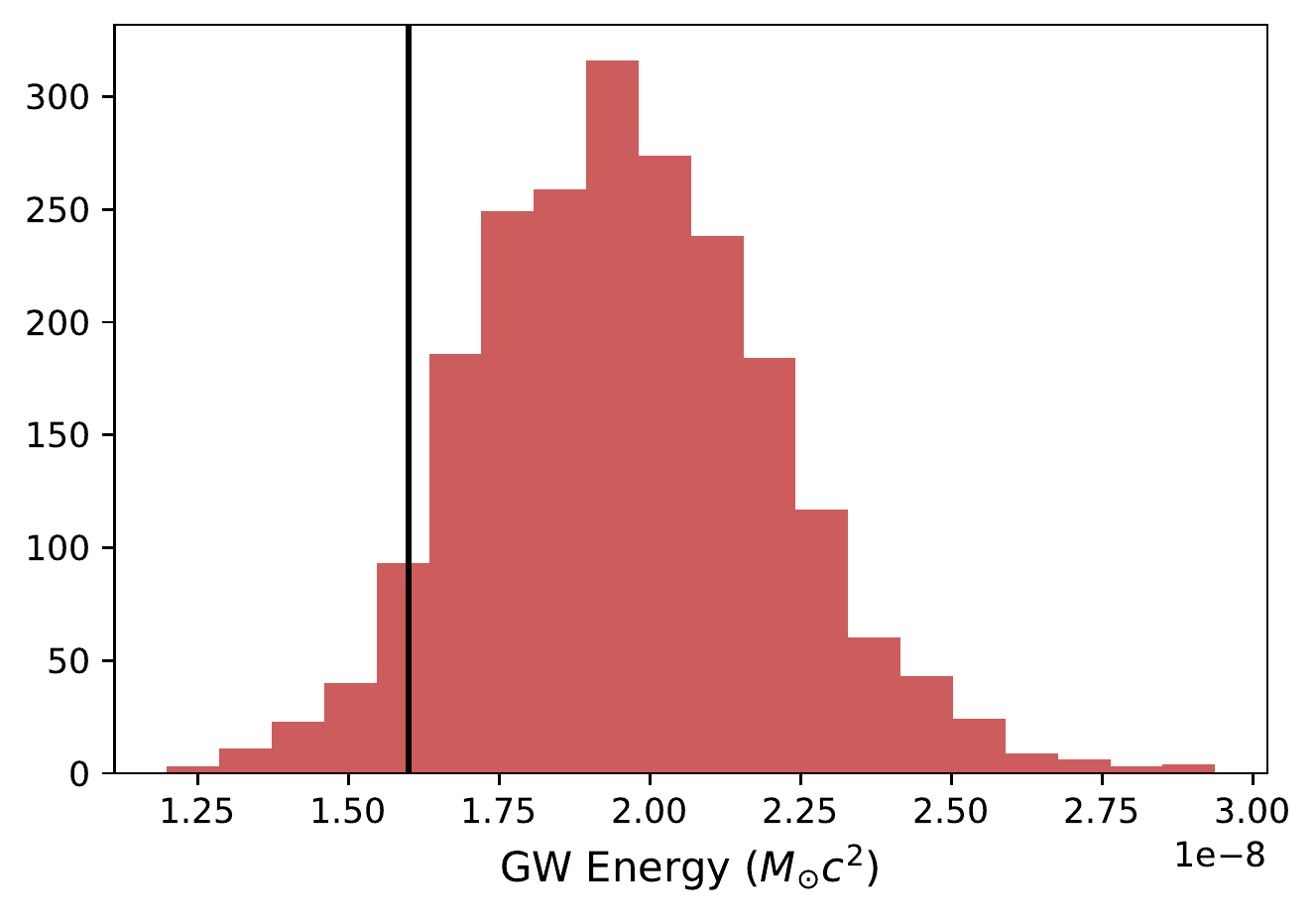}
\includegraphics[width=\columnwidth]{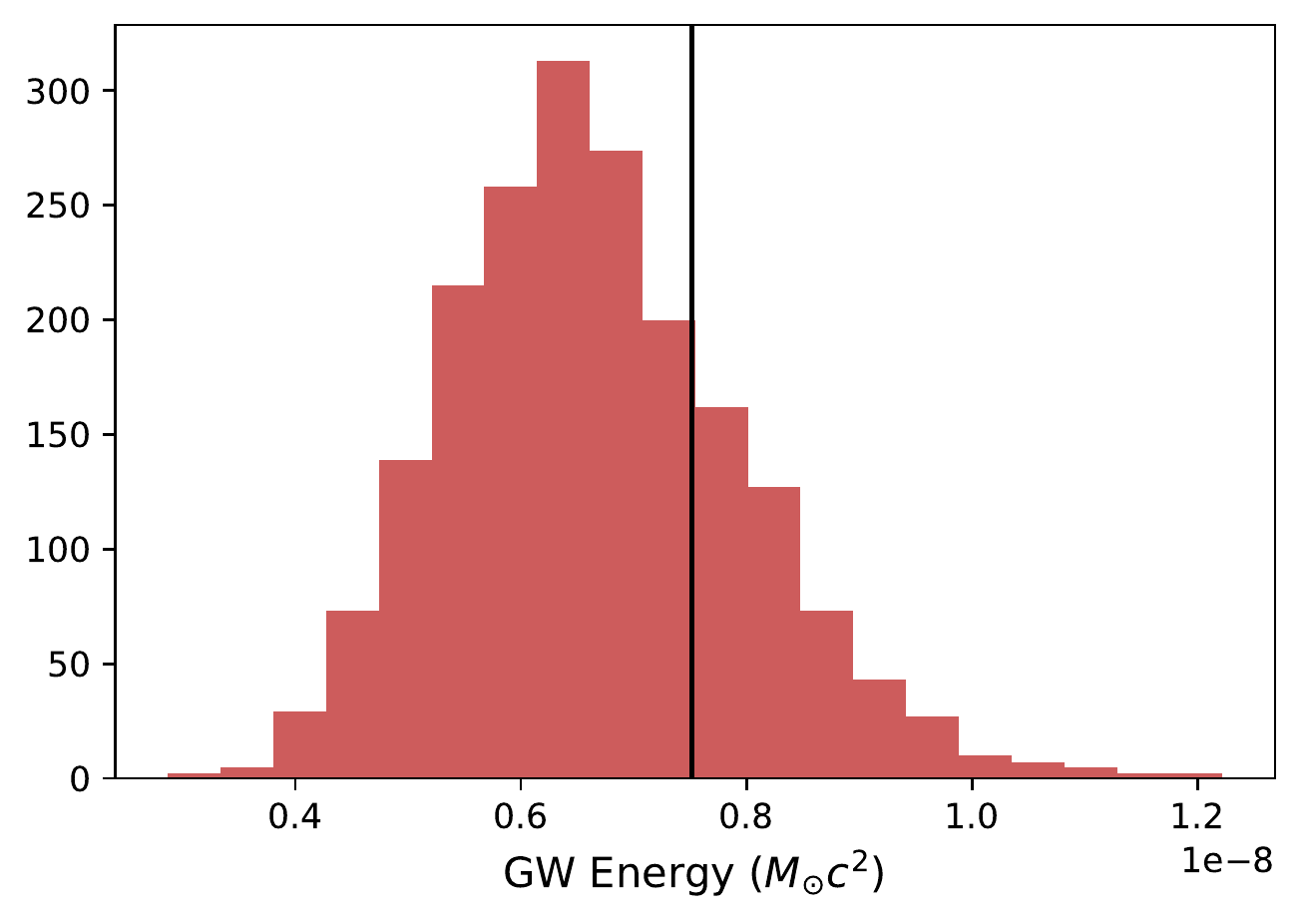}
\includegraphics[width=\columnwidth]{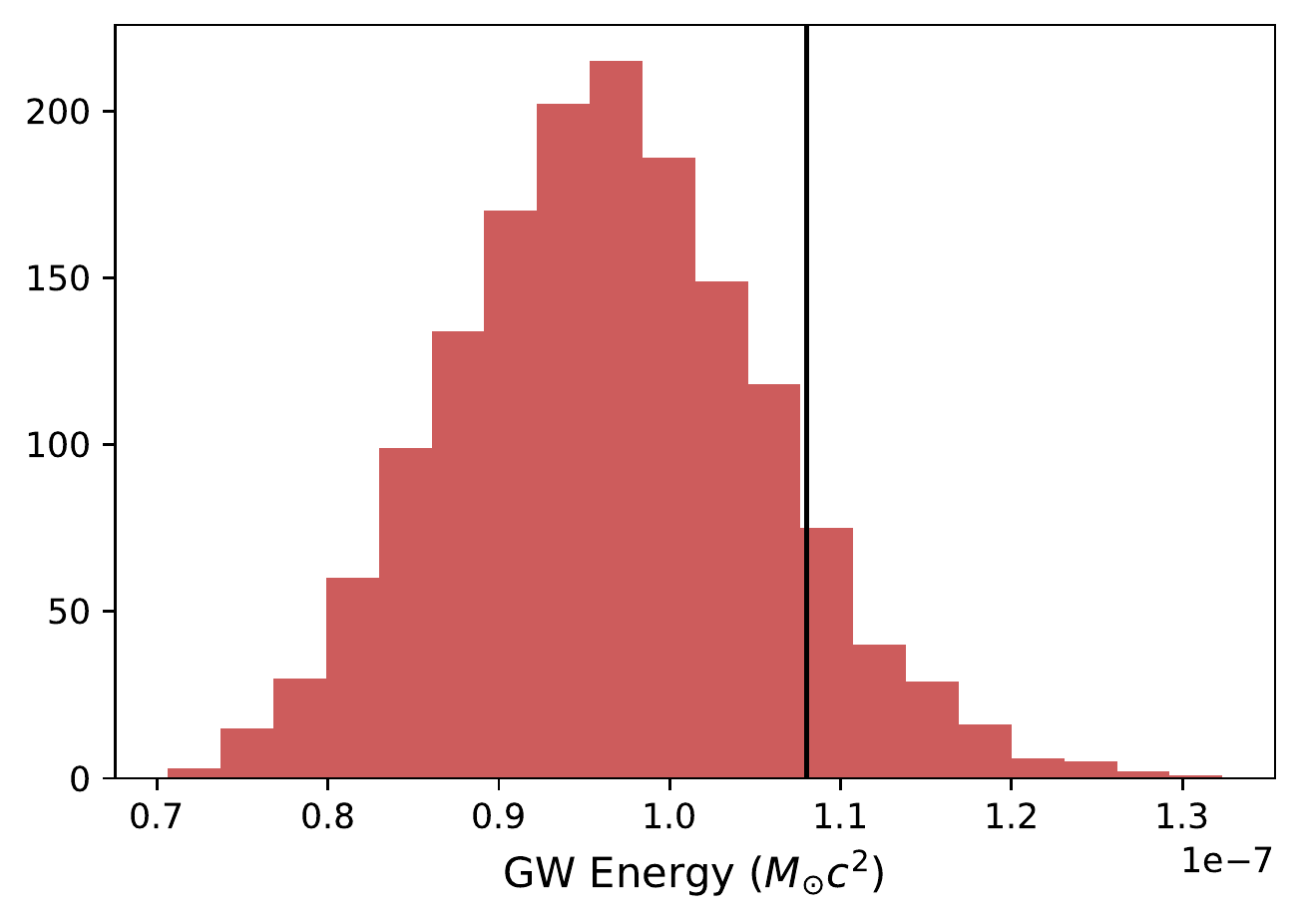}
\caption{The gravitational-wave energy measured by our analysis for model he3.5 (top left), model s18 (top right), model y20 (bottom left) and model m39 (bottom right) at a signal to noise ratio of 25. The black lines are the true values given by the CCSN waveform simulations. }
\label{fig:energy}
\end{figure*}

As the distance to a CCSN will likely be known, we can estimate the gravitational-wave energy. The gravitational-wave energy is given by, 
\begin{equation}
E_{GW} = \frac{h_\mathrm{rss}^2 \pi^2  c^3 d^2 f^2}{ G}  ,
\end{equation}
where $d$ is the distance, $f$ is the peak frequency, and $h_\mathrm{rss}$ is the root sum squared amplitude of the gravitational-wave signal. This equation is designed for narrow band signals, however it still works well for our supernova signals as the majority of their gravitational-wave energy occurs in a frequency band of a few hundred milli-seconds. This equation may not be appropriate for some other supernova models. We show the posterior distributions for the gravitational-wave energy in Figure \ref{fig:energy}. The posterior distributions are a good fit for the true values from our CCSN waveform simulations. Measuring the gravitational-wave energy may also inform us about other aspects of the explosion. Models with larger gravitational-wave energy tend to have larger explosion energies, therefore we may be able to put some constraint on the explosion energy by measuring the gravitational-wave energy. Further to this, Radice et al. \cite{2019ApJ...876L...9R} found a relationship between the total gravitational-wave energy and the time-integrated flux of turbulent kinetic energy $E_\mathrm{turb}$ from the gain region onto the PNS. This relationship is given by the equation,
\begin{equation}
E_{GW} = 3 \times 10^{43} \mathrm{erg} \left( \frac{E_\mathrm{turb}}{10^{50}\mathrm{erg}}  \right)^{1.88} ,
\end{equation}
which was further expanded in \cite{2019MNRAS.487.1178P} to include the time of shock revival $\tau$ and the Mach number $\mathrm{Ma}$ given by, 
%
\begin{multline}
E_{GW} = 4.2 \times 10^{43} \mathrm{erg} \left( \frac{f}{1000 \,\mathrm{Hz}} \right)^{2} \\  \left( \frac{\tau}{20\,\mathrm{ms}} \right)^{2}  \left( \frac{\mathrm{Ma}^{2}}{0.3} \right)  \left( \frac{E_\mathrm{turb}}{10^{50}\,\mathrm{erg}}  \right)^{2} .
\end{multline}
Therefore, using our measured frequency, shock revival time, and gravitational-wave energy we could make a statement about $E_\mathrm{turb}$ in the event of a real CCSN gravitational-wave detection. 

\section{Conclusions}
\label{sec:conclusion}

Ground based gravitational-wave detectors are currently preparing for their fourth observing runs. The upcoming increase in detector sensitivity may lead to the first discovery of gravitational waves from a CCSN. Understanding the relationship between the gravitational-wave signal and the source astrophysics will be essential for us to produce timely parameter estimation results in the event of a real CCSN gravitational-wave detection.  

In this paper, we add a new asymmetric chirplet model to the Bayesian parameter estimation code Bilby to represent the main emission mode visible in spectrograms of CCSN gravitational-wave signals. As CCSN signals have stochastic phase information in the time series, we also add a non-central chi-squared likelihood function so that we can carry out our analysis in the spectrogram domain, where the phase is no longer an issue, and the shape of the signal modes is most clearly visible. This likelihood function was used previously in \cite{2019PhRvD..99f3018R}, however they only use it to perform model selection, so this work is the first time that spectrograms have been used for CCSN parameter estimation. 

We find that when using the non-central chi-squared likelihood function we can determine that a signal is present at lower detector network SNRs than when the traditional Gaussian likelihood function and time series signal model is used. We show that a network SNR of 12 is required before we can start to constrain the signal parameters, and a network SNR of 16 is required before we can reconstruct the full duration of the signal mode. This will allow us to perform CCSN parameter estimation for sources at larger distances. Our model does not capture the stochastic fluctuations in signal amplitude, however they are not important for understanding the source astrophysics. Further to this, we show that performing the analysis in the spectrogram domain allows us to reconstruct a larger part of the signal mode than when the analysis is carried out in the time domain. 

Previous CCSN gravitational-wave data analysis studies have mainly focused on parameter estimation of the core-bounce signal, detection or model selection. Bizouard et al. also produce a reconstruction of the gravitational-wave signal g-mode. In this work we try and relate the parameters of our signal model directly to the astrophysical parameters of the source. We show how the signal frequency in our reconstructed spectrograms can be converted into posterior distributions on the mass and radius of the PNS. The best results are obtained at times when the gravitational-wave amplitude is highest. Errors at low frequencies are caused by a mis-match between the universal relations and our waveform simulations. At high frequencies, it becomes more difficult to distinguish the gravitational-wave signals from the noise, as the amplitude of the gravitational-wave signal decreases and the quality of the detector noise also decreases. This was not the case in the study by Bizouard et al., who use 2D supernova waveforms with a more consistent amplitude throughout the entire g-mode. Measuring a shorter duration of the full signal when the SNR is low means that we will only be able to predict the properties of the PNS for a shorter length of time.
The results may be improved further as work on universal relations between the source properties and gravitational-wave frequency improve. Understanding how the universal relations change when rotation or magnetic fields are included is currently not understood. Further to this, if we extend our analysis in future to include the low frequency SASI mode then we would also be able to use the same method to produce posterior distributions for the evolution of the shock radius. 

We show how the measured time of peak frequency is related to the shock revival time. For lower mass models, we find that the peak frequency occurs around the shock revival time, and for higher mass models the peak frequency occurs $\sim 100$\,ms later. It is likely that the progenitor mass and constraints on the core structure will already be known from electromagnetic or neutrino observations, which will help us constrain the shock revival time. Further to this, if the gravitational-wave signal has a low frequency SASI mode, that will also aid in determining the shock revival time as the SASI mode stops increasing in frequency with time once the shock has been revived. 

As the distance to a CCSN will be known, we can also measure the gravitational-wave energy. Simulations have shown that 
models with higher explosion energies tend to also have higher gravitational-wave energies.
However, the explosion energy from CCSN simulations still does not match the explosion energies observed electromagnetically. Recent studies have shown that the gravitational-wave energy is related to the turbulent kinetic energy from the gain regain onto the PNS. Therefore, using these relations, the posteriors we obtain for the gravitational-wave energy could be converted into measurements of the turbulent kinetic energy. 

In future work, we could expand our signal model to also include the core-bounce signal, which can be used to determine the rotation of the progenitor star. Including the low frequency SASI mode would allow us also to better constrain the shock revival time and also allow us to produce posterior distributions on the shock radius. In this work, we only include waveforms simulated with the CoCoNuT code. As there are some differences in results between different simulation codes, in future work we will need to test the robustness of our results by using CCSN waveforms from other simulation codes. It may also be possible to gain further knowledge of the source properties by doing a joint analysis with source information gained by electromagnetic or neutrino observations.

\begin{acknowledgments}
JP is supported by the Australian Research Council's (ARC) Discovery Early Career Researcher Award (DECRA) project number DE210101050, and the ARC Centre of Excellence for Gravitational Wave Discovery (OzGrav) project number CE170100004. BM acknowledges support by ARC Future Fellowship FT160100035. 
This work is based on simulations performed within computer time allocations from Astronomy Australia Limited's ASTAC scheme,  the National Computational Merit Allocation Scheme (NCMAS), and an Australasian Leadership computing grant on the NCI NF supercomputer Gadi. This research was supported by
resources provided by the Pawsey Supercomputing Centre, with funding from the Australian Government and the Government of Western Australia. 

\end{acknowledgments}


\bibliographystyle{apsrev}

\bibliography{bibfile}

\begin{thebibliography}{52}
\expandafter\ifx\csname natexlab\endcsname\relax\def\natexlab#1{#1}\fi
\expandafter\ifx\csname bibnamefont\endcsname\relax
  \def\bibnamefont#1{#1}\fi
\expandafter\ifx\csname bibfnamefont\endcsname\relax
  \def\bibfnamefont#1{#1}\fi
\expandafter\ifx\csname citenamefont\endcsname\relax
  \def\citenamefont#1{#1}\fi
\expandafter\ifx\csname url\endcsname\relax
  \def\url#1{\texttt{#1}}\fi
\expandafter\ifx\csname urlprefix\endcsname\relax\def\urlprefix{URL }\fi
\providecommand{\bibinfo}[2]{#2}
\providecommand{\eprint}[2][]{\url{#2}}

\bibitem[{\citenamefont{{Abbott} et~al.}(2019)\citenamefont{{Abbott}, {Abbott},
  {Abbott}, {Abraham}, {Acernese}, {Ackley}, {Adams}, {Adhikari}, {Adya},
  {Affeldt} et~al.}}]{2019PhRvX...9c1040A}
\bibinfo{author}{\bibfnamefont{B.~P.} \bibnamefont{{Abbott}}},
  \bibinfo{author}{\bibfnamefont{R.}~\bibnamefont{{Abbott}}},
  \bibinfo{author}{\bibfnamefont{T.~D.} \bibnamefont{{Abbott}}},
  \bibinfo{author}{\bibfnamefont{S.}~\bibnamefont{{Abraham}}},
  \bibinfo{author}{\bibfnamefont{F.}~\bibnamefont{{Acernese}}},
  \bibinfo{author}{\bibfnamefont{K.}~\bibnamefont{{Ackley}}},
  \bibinfo{author}{\bibfnamefont{C.}~\bibnamefont{{Adams}}},
  \bibinfo{author}{\bibfnamefont{R.~X.} \bibnamefont{{Adhikari}}},
  \bibinfo{author}{\bibfnamefont{V.~B.} \bibnamefont{{Adya}}},
  \bibinfo{author}{\bibfnamefont{C.}~\bibnamefont{{Affeldt}}},
  \bibnamefont{et~al.}, \bibinfo{journal}{Physical Review X}
  \textbf{\bibinfo{volume}{9}}, \bibinfo{eid}{031040} (\bibinfo{year}{2019}),
  \eprint{1811.12907}.

\bibitem[{\citenamefont{{Abbott} et~al.}(2021)\citenamefont{{Abbott}, {Abbott},
  {Abraham}, {Acernese}, {Ackley}, {Adams}, {Adams}, {Adhikari}, {Adya},
  {Affeldt} et~al.}}]{2021PhRvX..11b1053A}
\bibinfo{author}{\bibfnamefont{R.}~\bibnamefont{{Abbott}}},
  \bibinfo{author}{\bibfnamefont{T.~D.} \bibnamefont{{Abbott}}},
  \bibinfo{author}{\bibfnamefont{S.}~\bibnamefont{{Abraham}}},
  \bibinfo{author}{\bibfnamefont{F.}~\bibnamefont{{Acernese}}},
  \bibinfo{author}{\bibfnamefont{K.}~\bibnamefont{{Ackley}}},
  \bibinfo{author}{\bibfnamefont{A.}~\bibnamefont{{Adams}}},
  \bibinfo{author}{\bibfnamefont{C.}~\bibnamefont{{Adams}}},
  \bibinfo{author}{\bibfnamefont{R.~X.} \bibnamefont{{Adhikari}}},
  \bibinfo{author}{\bibfnamefont{V.~B.} \bibnamefont{{Adya}}},
  \bibinfo{author}{\bibfnamefont{C.}~\bibnamefont{{Affeldt}}},
  \bibnamefont{et~al.}, \bibinfo{journal}{Physical Review X}
  \textbf{\bibinfo{volume}{11}}, \bibinfo{eid}{021053} (\bibinfo{year}{2021}),
  \eprint{2010.14527}.

\bibitem[{\citenamefont{{The LIGO Scientific Collaboration}
  et~al.}(2021)\citenamefont{{The LIGO Scientific Collaboration}, {the Virgo
  Collaboration}, {the KAGRA Collaboration}, {Abbott}, {Abbott}, {Acernese},
  {Ackley}, {Adams}, {Adhikari}, {Adhikari} et~al.}}]{2021arXiv211103606T}
\bibinfo{author}{\bibnamefont{{The LIGO Scientific Collaboration}}},
  \bibinfo{author}{\bibnamefont{{the Virgo Collaboration}}},
  \bibinfo{author}{\bibnamefont{{the KAGRA Collaboration}}},
  \bibinfo{author}{\bibfnamefont{R.}~\bibnamefont{{Abbott}}},
  \bibinfo{author}{\bibfnamefont{T.~D.} \bibnamefont{{Abbott}}},
  \bibinfo{author}{\bibfnamefont{F.}~\bibnamefont{{Acernese}}},
  \bibinfo{author}{\bibfnamefont{K.}~\bibnamefont{{Ackley}}},
  \bibinfo{author}{\bibfnamefont{C.}~\bibnamefont{{Adams}}},
  \bibinfo{author}{\bibfnamefont{N.}~\bibnamefont{{Adhikari}}},
  \bibinfo{author}{\bibfnamefont{R.~X.} \bibnamefont{{Adhikari}}},
  \bibnamefont{et~al.}, \bibinfo{journal}{arXiv e-prints}
  \bibinfo{eid}{arXiv:2111.03606} (\bibinfo{year}{2021}), \eprint{2111.03606}.

\bibitem[{\citenamefont{{LIGO Scientific Collaboration}
  et~al.}(2015)\citenamefont{{LIGO Scientific Collaboration}, {Aasi}, {Abbott},
  {Abbott}, {Abbott}, {Abernathy}, {Ackley}, {Adams}, {Adams}, {Addesso}
  et~al.}}]{2015CQGra..32g4001L}
\bibinfo{author}{\bibnamefont{{LIGO Scientific Collaboration}}},
  \bibinfo{author}{\bibfnamefont{J.}~\bibnamefont{{Aasi}}},
  \bibinfo{author}{\bibfnamefont{B.~P.} \bibnamefont{{Abbott}}},
  \bibinfo{author}{\bibfnamefont{R.}~\bibnamefont{{Abbott}}},
  \bibinfo{author}{\bibfnamefont{T.}~\bibnamefont{{Abbott}}},
  \bibinfo{author}{\bibfnamefont{M.~R.} \bibnamefont{{Abernathy}}},
  \bibinfo{author}{\bibfnamefont{K.}~\bibnamefont{{Ackley}}},
  \bibinfo{author}{\bibfnamefont{C.}~\bibnamefont{{Adams}}},
  \bibinfo{author}{\bibfnamefont{T.}~\bibnamefont{{Adams}}},
  \bibinfo{author}{\bibfnamefont{P.}~\bibnamefont{{Addesso}}},
  \bibnamefont{et~al.}, \bibinfo{journal}{Classical and Quantum Gravity}
  \textbf{\bibinfo{volume}{32}}, \bibinfo{eid}{074001} (\bibinfo{year}{2015}),
  \eprint{1411.4547}.

\bibitem[{\citenamefont{{Acernese} et~al.}(2015)\citenamefont{{Acernese},
  {Agathos}, {Agatsuma}, {Aisa}, {Allemandou}, {Allocca}, {Amarni}, {Astone},
  {Balestri}, {Ballardin} et~al.}}]{2015CQGra..32b4001A}
\bibinfo{author}{\bibfnamefont{F.}~\bibnamefont{{Acernese}}},
  \bibinfo{author}{\bibfnamefont{M.}~\bibnamefont{{Agathos}}},
  \bibinfo{author}{\bibfnamefont{K.}~\bibnamefont{{Agatsuma}}},
  \bibinfo{author}{\bibfnamefont{D.}~\bibnamefont{{Aisa}}},
  \bibinfo{author}{\bibfnamefont{N.}~\bibnamefont{{Allemandou}}},
  \bibinfo{author}{\bibfnamefont{A.}~\bibnamefont{{Allocca}}},
  \bibinfo{author}{\bibfnamefont{J.}~\bibnamefont{{Amarni}}},
  \bibinfo{author}{\bibfnamefont{P.}~\bibnamefont{{Astone}}},
  \bibinfo{author}{\bibfnamefont{G.}~\bibnamefont{{Balestri}}},
  \bibinfo{author}{\bibfnamefont{G.}~\bibnamefont{{Ballardin}}},
  \bibnamefont{et~al.}, \bibinfo{journal}{Classical and Quantum Gravity}
  \textbf{\bibinfo{volume}{32}}, \bibinfo{eid}{024001} (\bibinfo{year}{2015}),
  \eprint{1408.3978}.

\bibitem[{\citenamefont{{Akutsu} et~al.}(2020)\citenamefont{{Akutsu}, {Ando},
  {Arai}, {Arai}, {Araki}, {Araya}, {Aritomi}, {Aso}, {Bae}, {Bae}
  et~al.}}]{2020arXiv200505574A}
\bibinfo{author}{\bibfnamefont{T.}~\bibnamefont{{Akutsu}}},
  \bibinfo{author}{\bibfnamefont{M.}~\bibnamefont{{Ando}}},
  \bibinfo{author}{\bibfnamefont{K.}~\bibnamefont{{Arai}}},
  \bibinfo{author}{\bibfnamefont{Y.}~\bibnamefont{{Arai}}},
  \bibinfo{author}{\bibfnamefont{S.}~\bibnamefont{{Araki}}},
  \bibinfo{author}{\bibfnamefont{A.}~\bibnamefont{{Araya}}},
  \bibinfo{author}{\bibfnamefont{N.}~\bibnamefont{{Aritomi}}},
  \bibinfo{author}{\bibfnamefont{Y.}~\bibnamefont{{Aso}}},
  \bibinfo{author}{\bibfnamefont{S.~W.} \bibnamefont{{Bae}}},
  \bibinfo{author}{\bibfnamefont{Y.~B.} \bibnamefont{{Bae}}},
  \bibnamefont{et~al.}, \bibinfo{journal}{arXiv e-prints}
  \bibinfo{eid}{arXiv:2005.05574} (\bibinfo{year}{2020}), \eprint{2005.05574}.

\bibitem[{\citenamefont{{Abbott} et~al.}(2020)\citenamefont{{Abbott}, {Abbott},
  {Abbott}, {Abraham}, {Acernese}, {Ackley}, {Adams}, {Adya}, {Affeldt},
  {Agathos} et~al.}}]{2020PhRvD.101h4002A}
\bibinfo{author}{\bibfnamefont{B.~P.} \bibnamefont{{Abbott}}},
  \bibinfo{author}{\bibfnamefont{R.}~\bibnamefont{{Abbott}}},
  \bibinfo{author}{\bibfnamefont{T.~D.} \bibnamefont{{Abbott}}},
  \bibinfo{author}{\bibfnamefont{S.}~\bibnamefont{{Abraham}}},
  \bibinfo{author}{\bibfnamefont{F.}~\bibnamefont{{Acernese}}},
  \bibinfo{author}{\bibfnamefont{K.}~\bibnamefont{{Ackley}}},
  \bibinfo{author}{\bibfnamefont{C.}~\bibnamefont{{Adams}}},
  \bibinfo{author}{\bibfnamefont{V.~B.} \bibnamefont{{Adya}}},
  \bibinfo{author}{\bibfnamefont{C.}~\bibnamefont{{Affeldt}}},
  \bibinfo{author}{\bibfnamefont{M.}~\bibnamefont{{Agathos}}},
  \bibnamefont{et~al.}, \bibinfo{journal}{\prd} \textbf{\bibinfo{volume}{101}},
  \bibinfo{eid}{084002} (\bibinfo{year}{2020}), \eprint{1908.03584}.

\bibitem[{\citenamefont{{Abdikamalov} et~al.}(2020)\citenamefont{{Abdikamalov},
  {Pagliaroli}, and {Radice}}}]{2020arXiv201004356A}
\bibinfo{author}{\bibfnamefont{E.}~\bibnamefont{{Abdikamalov}}},
  \bibinfo{author}{\bibfnamefont{G.}~\bibnamefont{{Pagliaroli}}},
  \bibnamefont{and} \bibinfo{author}{\bibfnamefont{D.}~\bibnamefont{{Radice}}},
  \bibinfo{journal}{arXiv e-prints} \bibinfo{eid}{arXiv:2010.04356}
  (\bibinfo{year}{2020}), \eprint{2010.04356}.

\bibitem[{\citenamefont{{Mezzacappa} et~al.}(2020)\citenamefont{{Mezzacappa},
  {Marronetti}, {Landfield}, {Lentz}, {Yakunin}, {Bruenn}, {Hix}, {Messer},
  {Endeve}, {Blondin} et~al.}}]{2020PhRvD.102b3027M}
\bibinfo{author}{\bibfnamefont{A.}~\bibnamefont{{Mezzacappa}}},
  \bibinfo{author}{\bibfnamefont{P.}~\bibnamefont{{Marronetti}}},
  \bibinfo{author}{\bibfnamefont{R.~E.} \bibnamefont{{Landfield}}},
  \bibinfo{author}{\bibfnamefont{E.~J.} \bibnamefont{{Lentz}}},
  \bibinfo{author}{\bibfnamefont{K.~N.} \bibnamefont{{Yakunin}}},
  \bibinfo{author}{\bibfnamefont{S.~W.} \bibnamefont{{Bruenn}}},
  \bibinfo{author}{\bibfnamefont{W.~R.} \bibnamefont{{Hix}}},
  \bibinfo{author}{\bibfnamefont{O.~E.~B.} \bibnamefont{{Messer}}},
  \bibinfo{author}{\bibfnamefont{E.}~\bibnamefont{{Endeve}}},
  \bibinfo{author}{\bibfnamefont{J.~M.} \bibnamefont{{Blondin}}},
  \bibnamefont{et~al.}, \bibinfo{journal}{\prd} \textbf{\bibinfo{volume}{102}},
  \bibinfo{eid}{023027} (\bibinfo{year}{2020}), \eprint{2007.15099}.

\bibitem[{\citenamefont{{Powell} and {M{\"u}ller}}(2020)}]{2020MNRAS.494.4665P}
\bibinfo{author}{\bibfnamefont{J.}~\bibnamefont{{Powell}}} \bibnamefont{and}
  \bibinfo{author}{\bibfnamefont{B.}~\bibnamefont{{M{\"u}ller}}},
  \bibinfo{journal}{\mnras} \textbf{\bibinfo{volume}{494}},
  \bibinfo{pages}{4665} (\bibinfo{year}{2020}), \eprint{2002.10115}.

\bibitem[{\citenamefont{{Andresen} et~al.}(2019)\citenamefont{{Andresen},
  {M{\"u}ller}, {Janka}, {Summa}, {Gill}, and {Zanolin}}}]{2019MNRAS.486.2238A}
\bibinfo{author}{\bibfnamefont{H.}~\bibnamefont{{Andresen}}},
  \bibinfo{author}{\bibfnamefont{E.}~\bibnamefont{{M{\"u}ller}}},
  \bibinfo{author}{\bibfnamefont{H.~T.} \bibnamefont{{Janka}}},
  \bibinfo{author}{\bibfnamefont{A.}~\bibnamefont{{Summa}}},
  \bibinfo{author}{\bibfnamefont{K.}~\bibnamefont{{Gill}}}, \bibnamefont{and}
  \bibinfo{author}{\bibfnamefont{M.}~\bibnamefont{{Zanolin}}},
  \bibinfo{journal}{\mnras} \textbf{\bibinfo{volume}{486}},
  \bibinfo{pages}{2238} (\bibinfo{year}{2019}), \eprint{1810.07638}.

\bibitem[{\citenamefont{{Powell} and {M{\"u}ller}}(2019)}]{2019MNRAS.487.1178P}
\bibinfo{author}{\bibfnamefont{J.}~\bibnamefont{{Powell}}} \bibnamefont{and}
  \bibinfo{author}{\bibfnamefont{B.}~\bibnamefont{{M{\"u}ller}}},
  \bibinfo{journal}{\mnras} \textbf{\bibinfo{volume}{487}},
  \bibinfo{pages}{1178} (\bibinfo{year}{2019}), \eprint{1812.05738}.

\bibitem[{\citenamefont{{Radice} et~al.}(2019)\citenamefont{{Radice},
  {Morozova}, {Burrows}, {Vartanyan}, and {Nagakura}}}]{2019ApJ...876L...9R}
\bibinfo{author}{\bibfnamefont{D.}~\bibnamefont{{Radice}}},
  \bibinfo{author}{\bibfnamefont{V.}~\bibnamefont{{Morozova}}},
  \bibinfo{author}{\bibfnamefont{A.}~\bibnamefont{{Burrows}}},
  \bibinfo{author}{\bibfnamefont{D.}~\bibnamefont{{Vartanyan}}},
  \bibnamefont{and}
  \bibinfo{author}{\bibfnamefont{H.}~\bibnamefont{{Nagakura}}},
  \bibinfo{journal}{\apjl} \textbf{\bibinfo{volume}{876}}, \bibinfo{eid}{L9}
  (\bibinfo{year}{2019}), \eprint{1812.07703}.

\bibitem[{\citenamefont{{O'Connor} and {Couch}}(2018)}]{2018ApJ...865...81O}
\bibinfo{author}{\bibfnamefont{E.~P.} \bibnamefont{{O'Connor}}}
  \bibnamefont{and} \bibinfo{author}{\bibfnamefont{S.~M.}
  \bibnamefont{{Couch}}}, \bibinfo{journal}{\apj}
  \textbf{\bibinfo{volume}{865}}, \bibinfo{eid}{81} (\bibinfo{year}{2018}),
  \eprint{1807.07579}.

\bibitem[{\citenamefont{{Pan} et~al.}(2018)\citenamefont{{Pan},
  {Liebend{\"o}rfer}, {Couch}, and {Thielemann}}}]{2018ApJ...857...13P}
\bibinfo{author}{\bibfnamefont{K.-C.} \bibnamefont{{Pan}}},
  \bibinfo{author}{\bibfnamefont{M.}~\bibnamefont{{Liebend{\"o}rfer}}},
  \bibinfo{author}{\bibfnamefont{S.~M.} \bibnamefont{{Couch}}},
  \bibnamefont{and} \bibinfo{author}{\bibfnamefont{F.-K.}
  \bibnamefont{{Thielemann}}}, \bibinfo{journal}{\apj}
  \textbf{\bibinfo{volume}{857}}, \bibinfo{eid}{13} (\bibinfo{year}{2018}),
  \eprint{1710.01690}.

\bibitem[{\citenamefont{{R{\"o}ver} et~al.}(2009)\citenamefont{{R{\"o}ver},
  {Bizouard}, {Christensen}, {Dimmelmeier}, {Heng}, and
  {Meyer}}}]{2009PhRvD..80j2004R}
\bibinfo{author}{\bibfnamefont{C.}~\bibnamefont{{R{\"o}ver}}},
  \bibinfo{author}{\bibfnamefont{M.-A.} \bibnamefont{{Bizouard}}},
  \bibinfo{author}{\bibfnamefont{N.}~\bibnamefont{{Christensen}}},
  \bibinfo{author}{\bibfnamefont{H.}~\bibnamefont{{Dimmelmeier}}},
  \bibinfo{author}{\bibfnamefont{I.~S.} \bibnamefont{{Heng}}},
  \bibnamefont{and} \bibinfo{author}{\bibfnamefont{R.}~\bibnamefont{{Meyer}}},
  \bibinfo{journal}{\prd} \textbf{\bibinfo{volume}{80}}, \bibinfo{eid}{102004}
  (\bibinfo{year}{2009}), \eprint{0909.1093}.

\bibitem[{\citenamefont{{Edwards}}(2021)}]{2021PhRvD.103b4025E}
\bibinfo{author}{\bibfnamefont{M.~C.} \bibnamefont{{Edwards}}},
  \bibinfo{journal}{\prd} \textbf{\bibinfo{volume}{103}}, \bibinfo{eid}{024025}
  (\bibinfo{year}{2021}), \eprint{2009.07367}.

\bibitem[{\citenamefont{{Abdikamalov} et~al.}(2014)\citenamefont{{Abdikamalov},
  {Gossan}, {DeMaio}, and {Ott}}}]{2014PhRvD..90d4001A}
\bibinfo{author}{\bibfnamefont{E.}~\bibnamefont{{Abdikamalov}}},
  \bibinfo{author}{\bibfnamefont{S.}~\bibnamefont{{Gossan}}},
  \bibinfo{author}{\bibfnamefont{A.~M.} \bibnamefont{{DeMaio}}},
  \bibnamefont{and} \bibinfo{author}{\bibfnamefont{C.~D.} \bibnamefont{{Ott}}},
  \bibinfo{journal}{\prd} \textbf{\bibinfo{volume}{90}}, \bibinfo{eid}{044001}
  (\bibinfo{year}{2014}), \eprint{1311.3678}.

\bibitem[{\citenamefont{{Richers} et~al.}(2017)\citenamefont{{Richers}, {Ott},
  {Abdikamalov}, {O'Connor}, and {Sullivan}}}]{2017PhRvD..95f3019R}
\bibinfo{author}{\bibfnamefont{S.}~\bibnamefont{{Richers}}},
  \bibinfo{author}{\bibfnamefont{C.~D.} \bibnamefont{{Ott}}},
  \bibinfo{author}{\bibfnamefont{E.}~\bibnamefont{{Abdikamalov}}},
  \bibinfo{author}{\bibfnamefont{E.}~\bibnamefont{{O'Connor}}},
  \bibnamefont{and}
  \bibinfo{author}{\bibfnamefont{C.}~\bibnamefont{{Sullivan}}},
  \bibinfo{journal}{\prd} \textbf{\bibinfo{volume}{95}}, \bibinfo{eid}{063019}
  (\bibinfo{year}{2017}), \eprint{1701.02752}.

\bibitem[{\citenamefont{{Edwards} et~al.}(2014)\citenamefont{{Edwards},
  {Meyer}, and {Christensen}}}]{2014InvPr..30k4008E}
\bibinfo{author}{\bibfnamefont{M.~C.} \bibnamefont{{Edwards}}},
  \bibinfo{author}{\bibfnamefont{R.}~\bibnamefont{{Meyer}}}, \bibnamefont{and}
  \bibinfo{author}{\bibfnamefont{N.}~\bibnamefont{{Christensen}}},
  \bibinfo{journal}{Inverse Problems} \textbf{\bibinfo{volume}{30}},
  \bibinfo{eid}{114008} (\bibinfo{year}{2014}), \eprint{1407.7549}.

\bibitem[{\citenamefont{{Engels} et~al.}(2014)\citenamefont{{Engels}, {Frey},
  and {Ott}}}]{2014PhRvD..90l4026E}
\bibinfo{author}{\bibfnamefont{W.~J.} \bibnamefont{{Engels}}},
  \bibinfo{author}{\bibfnamefont{R.}~\bibnamefont{{Frey}}}, \bibnamefont{and}
  \bibinfo{author}{\bibfnamefont{C.~D.} \bibnamefont{{Ott}}},
  \bibinfo{journal}{\prd} \textbf{\bibinfo{volume}{90}}, \bibinfo{eid}{124026}
  (\bibinfo{year}{2014}), \eprint{1406.1164}.

\bibitem[{\citenamefont{{Summerscales}
  et~al.}(2008)\citenamefont{{Summerscales}, {Burrows}, {Finn}, and
  {Ott}}}]{2008ApJ...678.1142S}
\bibinfo{author}{\bibfnamefont{T.~Z.} \bibnamefont{{Summerscales}}},
  \bibinfo{author}{\bibfnamefont{A.}~\bibnamefont{{Burrows}}},
  \bibinfo{author}{\bibfnamefont{L.~S.} \bibnamefont{{Finn}}},
  \bibnamefont{and} \bibinfo{author}{\bibfnamefont{C.~D.} \bibnamefont{{Ott}}},
  \bibinfo{journal}{\apj} \textbf{\bibinfo{volume}{678}}, \bibinfo{pages}{1142}
  (\bibinfo{year}{2008}), \eprint{0704.2157}.

\bibitem[{\citenamefont{Blondin et~al.}(2003)\citenamefont{Blondin, Mezzacappa,
  and DeMarino}}]{0004-637X-584-2-971}
\bibinfo{author}{\bibfnamefont{J.~M.} \bibnamefont{Blondin}},
  \bibinfo{author}{\bibfnamefont{A.}~\bibnamefont{Mezzacappa}},
  \bibnamefont{and} \bibinfo{author}{\bibfnamefont{C.}~\bibnamefont{DeMarino}},
  \bibinfo{journal}{The Astrophysical Journal} \textbf{\bibinfo{volume}{584}},
  \bibinfo{pages}{971} (\bibinfo{year}{2003}),
  \urlprefix\url{http://stacks.iop.org/0004-637X/584/i=2/a=971}.

\bibitem[{\citenamefont{{Blondin} and
  {Mezzacappa}}(2006)}]{2006ApJ...642..401B}
\bibinfo{author}{\bibfnamefont{J.~M.} \bibnamefont{{Blondin}}}
  \bibnamefont{and}
  \bibinfo{author}{\bibfnamefont{A.}~\bibnamefont{{Mezzacappa}}},
  \bibinfo{journal}{\apj} \textbf{\bibinfo{volume}{642}}, \bibinfo{pages}{401}
  (\bibinfo{year}{2006}), \eprint{astro-ph/0507181}.

\bibitem[{\citenamefont{{Foglizzo} et~al.}(2007)\citenamefont{{Foglizzo},
  {Galletti}, {Scheck}, and {Janka}}}]{2007ApJ...654.1006F}
\bibinfo{author}{\bibfnamefont{T.}~\bibnamefont{{Foglizzo}}},
  \bibinfo{author}{\bibfnamefont{P.}~\bibnamefont{{Galletti}}},
  \bibinfo{author}{\bibfnamefont{L.}~\bibnamefont{{Scheck}}}, \bibnamefont{and}
  \bibinfo{author}{\bibfnamefont{H.-T.} \bibnamefont{{Janka}}},
  \bibinfo{journal}{\apj} \textbf{\bibinfo{volume}{654}}, \bibinfo{pages}{1006}
  (\bibinfo{year}{2007}), \eprint{astro-ph/0606640}.

\bibitem[{\citenamefont{{Szczepanczyk}
  et~al.}(2021)\citenamefont{{Szczepanczyk}, {Antelis}, {Benjamin}, {Cavaglia},
  {Gondek-Rosinska}, {Hansen}, {Klimenko}, {Morales}, {Moreno}, {Mukherjee}
  et~al.}}]{2021arXiv210406462S}
\bibinfo{author}{\bibfnamefont{M.}~\bibnamefont{{Szczepanczyk}}},
  \bibinfo{author}{\bibfnamefont{J.}~\bibnamefont{{Antelis}}},
  \bibinfo{author}{\bibfnamefont{M.}~\bibnamefont{{Benjamin}}},
  \bibinfo{author}{\bibfnamefont{M.}~\bibnamefont{{Cavaglia}}},
  \bibinfo{author}{\bibfnamefont{D.}~\bibnamefont{{Gondek-Rosinska}}},
  \bibinfo{author}{\bibfnamefont{T.}~\bibnamefont{{Hansen}}},
  \bibinfo{author}{\bibfnamefont{S.}~\bibnamefont{{Klimenko}}},
  \bibinfo{author}{\bibfnamefont{M.}~\bibnamefont{{Morales}}},
  \bibinfo{author}{\bibfnamefont{C.}~\bibnamefont{{Moreno}}},
  \bibinfo{author}{\bibfnamefont{S.}~\bibnamefont{{Mukherjee}}},
  \bibnamefont{et~al.}, \bibinfo{journal}{arXiv e-prints}
  \bibinfo{eid}{arXiv:2104.06462} (\bibinfo{year}{2021}), \eprint{2104.06462}.

\bibitem[{\citenamefont{{Cornish} and
  {Littenberg}}(2015)}]{2015CQGra..32m5012C}
\bibinfo{author}{\bibfnamefont{N.~J.} \bibnamefont{{Cornish}}}
  \bibnamefont{and} \bibinfo{author}{\bibfnamefont{T.~B.}
  \bibnamefont{{Littenberg}}}, \bibinfo{journal}{Classical and Quantum Gravity}
  \textbf{\bibinfo{volume}{32}}, \bibinfo{eid}{135012} (\bibinfo{year}{2015}),
  \eprint{1410.3835}.

\bibitem[{\citenamefont{{Morozova} et~al.}(2018)\citenamefont{{Morozova},
  {Radice}, {Burrows}, and {Vartanyan}}}]{2018ApJ...861...10M}
\bibinfo{author}{\bibfnamefont{V.}~\bibnamefont{{Morozova}}},
  \bibinfo{author}{\bibfnamefont{D.}~\bibnamefont{{Radice}}},
  \bibinfo{author}{\bibfnamefont{A.}~\bibnamefont{{Burrows}}},
  \bibnamefont{and}
  \bibinfo{author}{\bibfnamefont{D.}~\bibnamefont{{Vartanyan}}},
  \bibinfo{journal}{\apj} \textbf{\bibinfo{volume}{861}}, \bibinfo{eid}{10}
  (\bibinfo{year}{2018}), \eprint{1801.01914}.

\bibitem[{\citenamefont{{Warren} et~al.}(2019)\citenamefont{{Warren}, {Couch},
  {O'Connor}, and {Morozova}}}]{2019arXiv191203328W}
\bibinfo{author}{\bibfnamefont{M.~L.} \bibnamefont{{Warren}}},
  \bibinfo{author}{\bibfnamefont{S.~M.} \bibnamefont{{Couch}}},
  \bibinfo{author}{\bibfnamefont{E.~P.} \bibnamefont{{O'Connor}}},
  \bibnamefont{and}
  \bibinfo{author}{\bibfnamefont{V.}~\bibnamefont{{Morozova}}},
  \bibinfo{journal}{arXiv e-prints} \bibinfo{eid}{arXiv:1912.03328}
  (\bibinfo{year}{2019}), \eprint{1912.03328}.

\bibitem[{\citenamefont{{Torres-Forn{\'e}}
  et~al.}(2019)\citenamefont{{Torres-Forn{\'e}}, {Cerd{\'a}-Dur{\'a}n},
  {Obergaulinger}, {M{\"u}ller}, and {Font}}}]{2019PhRvL.123e1102T}
\bibinfo{author}{\bibfnamefont{A.}~\bibnamefont{{Torres-Forn{\'e}}}},
  \bibinfo{author}{\bibfnamefont{P.}~\bibnamefont{{Cerd{\'a}-Dur{\'a}n}}},
  \bibinfo{author}{\bibfnamefont{M.}~\bibnamefont{{Obergaulinger}}},
  \bibinfo{author}{\bibfnamefont{B.}~\bibnamefont{{M{\"u}ller}}},
  \bibnamefont{and} \bibinfo{author}{\bibfnamefont{J.~A.}
  \bibnamefont{{Font}}}, \bibinfo{journal}{\prl}
  \textbf{\bibinfo{volume}{123}}, \bibinfo{eid}{051102} (\bibinfo{year}{2019}),
  \eprint{1902.10048}.

\bibitem[{\citenamefont{{Sotani} et~al.}(2021)\citenamefont{{Sotani},
  {Takiwaki}, and {Togashi}}}]{2021arXiv211003131S}
\bibinfo{author}{\bibfnamefont{H.}~\bibnamefont{{Sotani}}},
  \bibinfo{author}{\bibfnamefont{T.}~\bibnamefont{{Takiwaki}}},
  \bibnamefont{and}
  \bibinfo{author}{\bibfnamefont{H.}~\bibnamefont{{Togashi}}},
  \bibinfo{journal}{arXiv e-prints} \bibinfo{eid}{arXiv:2110.03131}
  (\bibinfo{year}{2021}), \eprint{2110.03131}.

\bibitem[{\citenamefont{{Bizouard} et~al.}(2021)\citenamefont{{Bizouard},
  {Maturana-Russel}, {Torres-Forn{\'e}}, {Obergaulinger},
  {Cerd{\'a}-Dur{\'a}n}, {Christensen}, {Font}, and
  {Meyer}}}]{2021PhRvD.103f3006B}
\bibinfo{author}{\bibfnamefont{M.-A.} \bibnamefont{{Bizouard}}},
  \bibinfo{author}{\bibfnamefont{P.}~\bibnamefont{{Maturana-Russel}}},
  \bibinfo{author}{\bibfnamefont{A.}~\bibnamefont{{Torres-Forn{\'e}}}},
  \bibinfo{author}{\bibfnamefont{M.}~\bibnamefont{{Obergaulinger}}},
  \bibinfo{author}{\bibfnamefont{P.}~\bibnamefont{{Cerd{\'a}-Dur{\'a}n}}},
  \bibinfo{author}{\bibfnamefont{N.}~\bibnamefont{{Christensen}}},
  \bibinfo{author}{\bibfnamefont{J.~A.} \bibnamefont{{Font}}},
  \bibnamefont{and} \bibinfo{author}{\bibfnamefont{R.}~\bibnamefont{{Meyer}}},
  \bibinfo{journal}{\prd} \textbf{\bibinfo{volume}{103}}, \bibinfo{eid}{063006}
  (\bibinfo{year}{2021}), \eprint{2012.00846}.

\bibitem[{\citenamefont{{Srivastava} et~al.}(2019)\citenamefont{{Srivastava},
  {Ballmer}, {Brown}, {Afle}, {Burrows}, {Radice}, and
  {Vartanyan}}}]{2019PhRvD.100d3026S}
\bibinfo{author}{\bibfnamefont{V.}~\bibnamefont{{Srivastava}}},
  \bibinfo{author}{\bibfnamefont{S.}~\bibnamefont{{Ballmer}}},
  \bibinfo{author}{\bibfnamefont{D.~A.} \bibnamefont{{Brown}}},
  \bibinfo{author}{\bibfnamefont{C.}~\bibnamefont{{Afle}}},
  \bibinfo{author}{\bibfnamefont{A.}~\bibnamefont{{Burrows}}},
  \bibinfo{author}{\bibfnamefont{D.}~\bibnamefont{{Radice}}}, \bibnamefont{and}
  \bibinfo{author}{\bibfnamefont{D.}~\bibnamefont{{Vartanyan}}},
  \bibinfo{journal}{\prd} \textbf{\bibinfo{volume}{100}}, \bibinfo{eid}{043026}
  (\bibinfo{year}{2019}), \eprint{1906.00084}.

\bibitem[{\citenamefont{{Heng}}(2009)}]{2009CQGra..26j5005H}
\bibinfo{author}{\bibfnamefont{I.~S.} \bibnamefont{{Heng}}},
  \bibinfo{journal}{Classical and Quantum Gravity}
  \textbf{\bibinfo{volume}{26}}, \bibinfo{eid}{105005} (\bibinfo{year}{2009}),
  \eprint{0810.5707}.

\bibitem[{\citenamefont{{Logue} et~al.}(2012)\citenamefont{{Logue}, {Ott},
  {Heng}, {Kalmus}, and {Scargill}}}]{2012PhRvD..86d4023L}
\bibinfo{author}{\bibfnamefont{J.}~\bibnamefont{{Logue}}},
  \bibinfo{author}{\bibfnamefont{C.~D.} \bibnamefont{{Ott}}},
  \bibinfo{author}{\bibfnamefont{I.~S.} \bibnamefont{{Heng}}},
  \bibinfo{author}{\bibfnamefont{P.}~\bibnamefont{{Kalmus}}}, \bibnamefont{and}
  \bibinfo{author}{\bibfnamefont{J.~H.~C.} \bibnamefont{{Scargill}}},
  \bibinfo{journal}{\prd} \textbf{\bibinfo{volume}{86}}, \bibinfo{eid}{044023}
  (\bibinfo{year}{2012}), \eprint{1202.3256}.

\bibitem[{\citenamefont{{Powell} et~al.}(2016)\citenamefont{{Powell}, {Gossan},
  {Logue}, and {Heng}}}]{2016PhRvD..94l3012P}
\bibinfo{author}{\bibfnamefont{J.}~\bibnamefont{{Powell}}},
  \bibinfo{author}{\bibfnamefont{S.~E.} \bibnamefont{{Gossan}}},
  \bibinfo{author}{\bibfnamefont{J.}~\bibnamefont{{Logue}}}, \bibnamefont{and}
  \bibinfo{author}{\bibfnamefont{I.~S.} \bibnamefont{{Heng}}},
  \bibinfo{journal}{\prd} \textbf{\bibinfo{volume}{94}}, \bibinfo{eid}{123012}
  (\bibinfo{year}{2016}), \eprint{1610.05573}.

\bibitem[{\citenamefont{{Roma} et~al.}(2019)\citenamefont{{Roma}, {Powell},
  {Heng}, and {Frey}}}]{2019PhRvD..99f3018R}
\bibinfo{author}{\bibfnamefont{V.}~\bibnamefont{{Roma}}},
  \bibinfo{author}{\bibfnamefont{J.}~\bibnamefont{{Powell}}},
  \bibinfo{author}{\bibfnamefont{I.~S.} \bibnamefont{{Heng}}},
  \bibnamefont{and} \bibinfo{author}{\bibfnamefont{R.}~\bibnamefont{{Frey}}},
  \bibinfo{journal}{\prd} \textbf{\bibinfo{volume}{99}}, \bibinfo{eid}{063018}
  (\bibinfo{year}{2019}), \eprint{1901.08692}.

\bibitem[{\citenamefont{{Coughlin} et~al.}(2014)\citenamefont{{Coughlin},
  {Christensen}, {Gair}, {Kandhasamy}, and {Thrane}}}]{2014CQGra..31p5012C}
\bibinfo{author}{\bibfnamefont{M.}~\bibnamefont{{Coughlin}}},
  \bibinfo{author}{\bibfnamefont{N.}~\bibnamefont{{Christensen}}},
  \bibinfo{author}{\bibfnamefont{J.}~\bibnamefont{{Gair}}},
  \bibinfo{author}{\bibfnamefont{S.}~\bibnamefont{{Kandhasamy}}},
  \bibnamefont{and} \bibinfo{author}{\bibfnamefont{E.}~\bibnamefont{{Thrane}}},
  \bibinfo{journal}{Classical and Quantum Gravity}
  \textbf{\bibinfo{volume}{31}}, \bibinfo{eid}{165012} (\bibinfo{year}{2014}),
  \eprint{1404.4642}.

\bibitem[{\citenamefont{{Astone} et~al.}(2018)\citenamefont{{Astone},
  {Cerd{\'a}-Dur{\'a}n}, {Di Palma}, {Drago}, {Muciaccia}, {Palomba}, and
  {Ricci}}}]{2018PhRvD..98l2002A}
\bibinfo{author}{\bibfnamefont{P.}~\bibnamefont{{Astone}}},
  \bibinfo{author}{\bibfnamefont{P.}~\bibnamefont{{Cerd{\'a}-Dur{\'a}n}}},
  \bibinfo{author}{\bibfnamefont{I.}~\bibnamefont{{Di Palma}}},
  \bibinfo{author}{\bibfnamefont{M.}~\bibnamefont{{Drago}}},
  \bibinfo{author}{\bibfnamefont{F.}~\bibnamefont{{Muciaccia}}},
  \bibinfo{author}{\bibfnamefont{C.}~\bibnamefont{{Palomba}}},
  \bibnamefont{and} \bibinfo{author}{\bibfnamefont{F.}~\bibnamefont{{Ricci}}},
  \bibinfo{journal}{\prd} \textbf{\bibinfo{volume}{98}}, \bibinfo{eid}{122002}
  (\bibinfo{year}{2018}), \eprint{1812.05363}.

\bibitem[{\citenamefont{{L{\'o}pez} et~al.}(2021)\citenamefont{{L{\'o}pez}, {Di
  Palma}, {Drago}, {Cerd{\'a}-Dur{\'a}n}, and {Ricci}}}]{2021PhRvD.103f3011L}
\bibinfo{author}{\bibfnamefont{M.}~\bibnamefont{{L{\'o}pez}}},
  \bibinfo{author}{\bibfnamefont{I.}~\bibnamefont{{Di Palma}}},
  \bibinfo{author}{\bibfnamefont{M.}~\bibnamefont{{Drago}}},
  \bibinfo{author}{\bibfnamefont{P.}~\bibnamefont{{Cerd{\'a}-Dur{\'a}n}}},
  \bibnamefont{and} \bibinfo{author}{\bibfnamefont{F.}~\bibnamefont{{Ricci}}},
  \bibinfo{journal}{\prd} \textbf{\bibinfo{volume}{103}}, \bibinfo{eid}{063011}
  (\bibinfo{year}{2021}).

\bibitem[{\citenamefont{{Ashton} et~al.}(2019)\citenamefont{{Ashton},
  {H{\"u}bner}, {Lasky}, {Talbot}, {Ackley}, {Biscoveanu}, {Chu}, {Divakarla},
  {Easter}, {Goncharov} et~al.}}]{2019ApJS..241...27A}
\bibinfo{author}{\bibfnamefont{G.}~\bibnamefont{{Ashton}}},
  \bibinfo{author}{\bibfnamefont{M.}~\bibnamefont{{H{\"u}bner}}},
  \bibinfo{author}{\bibfnamefont{P.~D.} \bibnamefont{{Lasky}}},
  \bibinfo{author}{\bibfnamefont{C.}~\bibnamefont{{Talbot}}},
  \bibinfo{author}{\bibfnamefont{K.}~\bibnamefont{{Ackley}}},
  \bibinfo{author}{\bibfnamefont{S.}~\bibnamefont{{Biscoveanu}}},
  \bibinfo{author}{\bibfnamefont{Q.}~\bibnamefont{{Chu}}},
  \bibinfo{author}{\bibfnamefont{A.}~\bibnamefont{{Divakarla}}},
  \bibinfo{author}{\bibfnamefont{P.~J.} \bibnamefont{{Easter}}},
  \bibinfo{author}{\bibfnamefont{B.}~\bibnamefont{{Goncharov}}},
  \bibnamefont{et~al.}, \bibinfo{journal}{\apjs}
  \textbf{\bibinfo{volume}{241}}, \bibinfo{eid}{27} (\bibinfo{year}{2019}),
  \eprint{1811.02042}.

\bibitem[{\citenamefont{{Dimmelmeier} et~al.}(2002)\citenamefont{{Dimmelmeier},
  {Font}, and {M{\"u}ller}}}]{2002A&A...393..523D}
\bibinfo{author}{\bibfnamefont{H.}~\bibnamefont{{Dimmelmeier}}},
  \bibinfo{author}{\bibfnamefont{J.~A.} \bibnamefont{{Font}}},
  \bibnamefont{and}
  \bibinfo{author}{\bibfnamefont{E.}~\bibnamefont{{M{\"u}ller}}},
  \bibinfo{journal}{\aap} \textbf{\bibinfo{volume}{393}}, \bibinfo{pages}{523}
  (\bibinfo{year}{2002}), \eprint{astro-ph/0204289}.

\bibitem[{\citenamefont{{M{\"u}ller} et~al.}(2010)\citenamefont{{M{\"u}ller},
  {Janka}, and {Dimmelmeier}}}]{2010ApJS..189..104M}
\bibinfo{author}{\bibfnamefont{B.}~\bibnamefont{{M{\"u}ller}}},
  \bibinfo{author}{\bibfnamefont{H.-T.} \bibnamefont{{Janka}}},
  \bibnamefont{and}
  \bibinfo{author}{\bibfnamefont{H.}~\bibnamefont{{Dimmelmeier}}},
  \bibinfo{journal}{\apjs} \textbf{\bibinfo{volume}{189}}, \bibinfo{pages}{104}
  (\bibinfo{year}{2010}), \eprint{1001.4841}.

\bibitem[{\citenamefont{{Aguilera-Dena}
  et~al.}(2018)\citenamefont{{Aguilera-Dena}, {Langer}, {Moriya}, and
  {Schootemeijer}}}]{aguilera_18}
\bibinfo{author}{\bibfnamefont{D.~R.} \bibnamefont{{Aguilera-Dena}}},
  \bibinfo{author}{\bibfnamefont{N.}~\bibnamefont{{Langer}}},
  \bibinfo{author}{\bibfnamefont{T.~J.} \bibnamefont{{Moriya}}},
  \bibnamefont{and}
  \bibinfo{author}{\bibfnamefont{A.}~\bibnamefont{{Schootemeijer}}},
  \bibinfo{journal}{\apj} \textbf{\bibinfo{volume}{858}}, \bibinfo{eid}{115}
  (\bibinfo{year}{2018}), \eprint{1804.07317}.

\bibitem[{\citenamefont{{Yoon}}(2017)}]{yoon_17}
\bibinfo{author}{\bibfnamefont{S.-C.} \bibnamefont{{Yoon}}},
  \bibinfo{journal}{\mnras} \textbf{\bibinfo{volume}{470}},
  \bibinfo{pages}{3970} (\bibinfo{year}{2017}), \eprint{1706.04716}.

\bibitem[{\citenamefont{{Tauris} et~al.}(2015)\citenamefont{{Tauris}, {Langer},
  and {Podsiadlowski}}}]{tauris_15}
\bibinfo{author}{\bibfnamefont{T.~M.} \bibnamefont{{Tauris}}},
  \bibinfo{author}{\bibfnamefont{N.}~\bibnamefont{{Langer}}}, \bibnamefont{and}
  \bibinfo{author}{\bibfnamefont{P.}~\bibnamefont{{Podsiadlowski}}},
  \bibinfo{journal}{\mnras} \textbf{\bibinfo{volume}{451}},
  \bibinfo{pages}{2123} (\bibinfo{year}{2015}), \eprint{1505.00270}.

\bibitem[{\citenamefont{{Romero-Shaw} et~al.}(2020)\citenamefont{{Romero-Shaw},
  {Talbot}, {Biscoveanu}, {D'Emilio}, {Ashton}, {Berry}, {Coughlin},
  {Galaudage}, {Hoy}, {H{\"u}bner} et~al.}}]{2020MNRAS.499.3295R}
\bibinfo{author}{\bibfnamefont{I.~M.} \bibnamefont{{Romero-Shaw}}},
  \bibinfo{author}{\bibfnamefont{C.}~\bibnamefont{{Talbot}}},
  \bibinfo{author}{\bibfnamefont{S.}~\bibnamefont{{Biscoveanu}}},
  \bibinfo{author}{\bibfnamefont{V.}~\bibnamefont{{D'Emilio}}},
  \bibinfo{author}{\bibfnamefont{G.}~\bibnamefont{{Ashton}}},
  \bibinfo{author}{\bibfnamefont{C.~P.~L.} \bibnamefont{{Berry}}},
  \bibinfo{author}{\bibfnamefont{S.}~\bibnamefont{{Coughlin}}},
  \bibinfo{author}{\bibfnamefont{S.}~\bibnamefont{{Galaudage}}},
  \bibinfo{author}{\bibfnamefont{C.}~\bibnamefont{{Hoy}}},
  \bibinfo{author}{\bibfnamefont{M.}~\bibnamefont{{H{\"u}bner}}},
  \bibnamefont{et~al.}, \bibinfo{journal}{\mnras}
  \textbf{\bibinfo{volume}{499}}, \bibinfo{pages}{3295} (\bibinfo{year}{2020}),
  \eprint{2006.00714}.

\bibitem[{\citenamefont{Finn}(1992)}]{PhysRevD.46.5236}
\bibinfo{author}{\bibfnamefont{L.~S.} \bibnamefont{Finn}},
  \bibinfo{journal}{Phys. Rev. D} \textbf{\bibinfo{volume}{46}},
  \bibinfo{pages}{5236} (\bibinfo{year}{1992}),
  \urlprefix\url{https://link.aps.org/doi/10.1103/PhysRevD.46.5236}.

\bibitem[{\citenamefont{Demirli and Saniie}(2014)}]{DEMIRLI2014907}
\bibinfo{author}{\bibfnamefont{R.}~\bibnamefont{Demirli}} \bibnamefont{and}
  \bibinfo{author}{\bibfnamefont{J.}~\bibnamefont{Saniie}},
  \bibinfo{journal}{Journal of the Franklin Institute}
  \textbf{\bibinfo{volume}{351}}, \bibinfo{pages}{907} (\bibinfo{year}{2014}),
  ISSN \bibinfo{issn}{0016-0032},
  \urlprefix\url{https://www.sciencedirect.com/science/article/pii/S0016003213003657}.

\bibitem[{\citenamefont{{Drago} et~al.}(2020)\citenamefont{{Drago}, {Gayathri},
  {Klimenko}, {Lazzaro}, {Milotti}, {Mitselmakher}, {Necula}, {O'Brian},
  {Prodi}, {Salemi} et~al.}}]{2020arXiv200612604D}
\bibinfo{author}{\bibfnamefont{M.}~\bibnamefont{{Drago}}},
  \bibinfo{author}{\bibfnamefont{V.}~\bibnamefont{{Gayathri}}},
  \bibinfo{author}{\bibfnamefont{S.}~\bibnamefont{{Klimenko}}},
  \bibinfo{author}{\bibfnamefont{C.}~\bibnamefont{{Lazzaro}}},
  \bibinfo{author}{\bibfnamefont{E.}~\bibnamefont{{Milotti}}},
  \bibinfo{author}{\bibfnamefont{G.}~\bibnamefont{{Mitselmakher}}},
  \bibinfo{author}{\bibfnamefont{V.}~\bibnamefont{{Necula}}},
  \bibinfo{author}{\bibfnamefont{B.}~\bibnamefont{{O'Brian}}},
  \bibinfo{author}{\bibfnamefont{G.~A.} \bibnamefont{{Prodi}}},
  \bibinfo{author}{\bibfnamefont{F.}~\bibnamefont{{Salemi}}},
  \bibnamefont{et~al.}, \bibinfo{journal}{arXiv e-prints}
  \bibinfo{eid}{arXiv:2006.12604} (\bibinfo{year}{2020}), \eprint{2006.12604}.

\bibitem[{\citenamefont{{Lynch} et~al.}(2015)\citenamefont{{Lynch}, {Vitale},
  {Essick}, {Katsavounidis}, and {Robinet}}}]{2015arXiv151105955L}
\bibinfo{author}{\bibfnamefont{R.}~\bibnamefont{{Lynch}}},
  \bibinfo{author}{\bibfnamefont{S.}~\bibnamefont{{Vitale}}},
  \bibinfo{author}{\bibfnamefont{R.}~\bibnamefont{{Essick}}},
  \bibinfo{author}{\bibfnamefont{E.}~\bibnamefont{{Katsavounidis}}},
  \bibnamefont{and}
  \bibinfo{author}{\bibfnamefont{F.}~\bibnamefont{{Robinet}}},
  \bibinfo{journal}{arXiv e-prints} \bibinfo{eid}{arXiv:1511.05955}
  (\bibinfo{year}{2015}), \eprint{1511.05955}.

\bibitem[{\citenamefont{{Veitch} et~al.}(2015)\citenamefont{{Veitch},
  {Raymond}, {Farr}, {Farr}, {Graff}, {Vitale}, {Aylott}, {Blackburn},
  {Christensen}, {Coughlin} et~al.}}]{2015PhRvD..91d2003V}
\bibinfo{author}{\bibfnamefont{J.}~\bibnamefont{{Veitch}}},
  \bibinfo{author}{\bibfnamefont{V.}~\bibnamefont{{Raymond}}},
  \bibinfo{author}{\bibfnamefont{B.}~\bibnamefont{{Farr}}},
  \bibinfo{author}{\bibfnamefont{W.}~\bibnamefont{{Farr}}},
  \bibinfo{author}{\bibfnamefont{P.}~\bibnamefont{{Graff}}},
  \bibinfo{author}{\bibfnamefont{S.}~\bibnamefont{{Vitale}}},
  \bibinfo{author}{\bibfnamefont{B.}~\bibnamefont{{Aylott}}},
  \bibinfo{author}{\bibfnamefont{K.}~\bibnamefont{{Blackburn}}},
  \bibinfo{author}{\bibfnamefont{N.}~\bibnamefont{{Christensen}}},
  \bibinfo{author}{\bibfnamefont{M.}~\bibnamefont{{Coughlin}}},
  \bibnamefont{et~al.}, \bibinfo{journal}{\prd} \textbf{\bibinfo{volume}{91}},
  \bibinfo{eid}{042003} (\bibinfo{year}{2015}), \eprint{1409.7215}.

\end{thebibliography}

\end{document}